\documentclass{ws-rv975x65}
\usepackage{ws-rv-van}

\usepackage{bm}
\usepackage[usenames]{color}

\makeindex

\def\bea{\begin{eqnarray}}
\def\eea{\end{eqnarray}}
\def\bphi{\bm{\phi}}
\def\bsigma{\bm{\sigma}}
\def\deepstrut{\vrule height 1.5ex depth 2.5ex width 0pt}
\def\rmd{\text{d}}
\def\rme{\text{e}}
\def\bJ{\bm{J}}
\def\bx{\bm{x}}
\def\bphi{\bm{\phi}}
\def\vecr{\bm{\hat{r}}}
\def\bomega{\bm{\omega}}

\begin{document}

\begin{center} {\large {\bf 
Review of Rotational Symmetry Breaking \\in Baby Skyrme Models\footnote{to appear in: G. Brown and M. Rho, Eds., {\it Multifaceted Skyrmions}, (World Scientific, Singapore, 2009).}
\footnote{A version of this manuscript with higher-resolution figures is available at \url{www.tau.ac.il/~itayhe/SkReview/SkReview.rar}.}
}} \vspace{16pt} \end{center}

\author[M. Karliner and I. Hen]{Marek Karliner and Itay Hen}

\address{Raymond and Beverly Sackler School of Physics and Astronomy \\
Tel-Aviv University, Tel-Aviv 69978, Israel.\\
marek@proton.tau.ac.il}

\begin{abstract}
We discuss one of the most
interesting phenomena exhibited by baby skyrmions -- 
breaking of rotational symmetry. 
The topics we will deal with here include the appearance of rotational symmetry breaking 
in the static solutions of baby Skyrme models, both in flat as well as in curved spaces,
the zero-temperature crystalline structure of baby skyrmions, and 
finally, the appearance of 
spontaneous breaking of rotational symmetry in rotating baby skyrmions.
\end{abstract}

\body

\setcounter{chapter}{1}

\section{Breaking of Rotational Symmetry in Baby Skyrme Models}\label{sec1.1}
The Skyrme model \cite{Skyrme1,Skyrme2} is an SU(2)-valued nonlinear theory for pions in (3+1) dimensions
with topological soliton solutions called skyrmions. 
Apart from a kinetic term, the Lagrangian of the model contains 
a `Skyrme' term which is of the fourth order in derivatives, 
and is used to introduce scale to the model \cite{3DSkyrme}. 
The existence of stable solutions in the Skyrme model is a consequence of the nontrivial
topology of the mapping $\mathcal{M}$ of the physical space
into the field space  at a given time,
$\quad \mathcal{M}:  S^3 \to SU(2) \cong S^3,\quad$ where 
the physical space $\mathbb{R}^3$  is compactified to $S^3$
by requiring the spatial infinity to be equivalent in
each direction.
The topology which stems from this one-point 
compactification allows the classification of maps into equivalence classes,
each of which has a unique conserved quantity called the topological charge.
\par
The Skyrme model has an analogue in (2+1) dimensions known as the baby Skyrme model,
also admitting stable field configurations of a solitonic nature \cite{Old1}.
Due to its lower dimension, the baby Skyrme model serves as a simplification of the original model,
but nonetheless it has a physical significance in its own right,
having several applications in condensed-matter physics \cite{Condensed},
specifically in ferromagnetic quantum Hall systems \cite{QHE,QHF1,QHF2,QHF3}.
There, baby skyrmions describe the excitations relative to 
ferromagnetic quantum Hall states, 
in terms of a gradient expansion in the spin density, a field with properties analogous to
the pion field in the $3$D case \cite{Lee}.

The target manifold in the baby model is described by a three-dimensional vector 
$\bphi=(\phi_1,\phi_2,\phi_3)$  with the constraint $\bphi \cdot \bphi=1$. 
In analogy with the $(3+1)$D case,
the domain of this model $\mathbb{R}^2$ is compactified to $S^2$,
yielding the topology required
for the classification of its field configurations into classes with conserved topological charges.
The Lagrangian density of the baby Skyrme model is given by:
\bea \label{eq:BabyLag}
\mathcal{L}=\frac1{2} \partial_{\mu} \bphi \cdot \partial^{\mu} \bphi
- \frac{\kappa^2}{2}\big[(\partial_{\mu} \bphi \cdot \partial^{\mu} \bphi)^2 -
(\partial_{\mu} \bphi \cdot \partial_{\nu} \bphi) 
\cdot (\partial^{\mu} \bphi \cdot \partial^{\nu} \bphi)
\big]
-U(\phi_3) \,,\nonumber \\
\eea
and consists of a kinetic term, a Skyrme term and a potential term. 
\par
While in (3+1) dimensions the latter term is optional \cite{TopoSol},
its presence in the (2+1)D model is necessary
for the stability of the solutions.
However, aside from the requirement that the potential vanishes at infinity for a given vacuum field
value (normally taken to be $\bphi^{(0)}=(0,0,1)$), 
its exact form is arbitrary and gives rise to a rich family
of possible baby-Skyrme models, several of which have been studied in  detail in the literature.
The simplest potential is the `holomorphic' model with
$U(\phi_3)=\mu^2 (1-\phi_3)^4$ \cite{Holo1,Holo2,Holo3}.
It is known to have a stable solution only in the charge-one sector (the name 
refers to the fact that the stable solution has an analytic form in terms of holomorphic functions). 
The model with the potential
$U(\phi_3)=\mu^2 (1-\phi_3)$ (commonly referred to as the `old' model) 
has also been extensively studied.
This potential gives rise to very structured non-rotationally-symmetric 
multi-skyrmions \cite{Old1,Old2}.
Another model with $U(\phi_3)=\mu^2 (1-\phi_3^2)$ produces ring-like multi-skyrmions \cite{Weidig}.
Other double-vacuum potentials which give rise to other types of solutions
have also been studied \cite{babyClass}. 
\par
Clearly, the form of the potential term
has a decisive effect on the properties of the minimal energy configurations
of the model.
It is then worthwhile to see how the multisolitons of the baby Skyrme model look like 
for the one-parametric 
family of potentials $U=\mu^2 (1-\phi_3)^s$ which generalizes
the `old` model ($s=1$) and the holomorphic model ($s=4$) \cite{HK1}.
As it turns out, the value of the parameter $s$ 
has dramatic effects on the static solutions
of the model, both quantitatively and qualitatively, in the sense
that it can be viewed as a `control' parameter responsible
for the repulsion or attraction between skyrmions, which in turn determines 
whether or not the minimal-energy configuration breaks rotational symmetry.

The Lagrangian density is now:
\bea \label{eq:BabyLagFamily}
 \mathcal{L}=\frac1{2} \partial_{\mu} \bphi \cdot \partial^{\mu} \bphi
- \frac{\kappa^2}{2}\left( (\partial_{\mu} \bphi \cdot \partial^{\mu} \bphi)^2 -
(\partial_{\mu} \bphi \cdot \partial_{\nu} \bphi) 
\cdot (\partial^{\mu} \bphi \cdot \partial^{\nu} \bphi)
\right)
-\mu^2 (1-\phi_3)^s \,,
\nonumber \\
\eea
and contains three free parameters, namely $\kappa, \mu$ and $s$.
Since either $\kappa$ or $\mu$ may be scaled away,
the parameter space of this model is in fact only two dimensional.
Our main goal here is to study the effects of these parameters
on the static solutions of the model within each topological sector.  \par
The multi-skyrmions of our model
are those field configurations which minimize the static energy
functional within each topological sector. 
In polar coordinates the energy functional is given by
\bea \label{eq:O3energy}
 E=\int r \, \rmd r \, \rmd \theta \left(
\frac1{2} (\partial_r \bphi \cdot \partial_r \bphi 
+ \frac1{r^2}\partial_{\theta} \bphi \cdot \partial_{\theta} \bphi)
+\frac{\kappa^2}{2} \frac{(\partial_r \bphi \times \partial_{\theta} \bphi)^2}{r^2}
+\mu^2 (1-\phi_3)^s
\right) \;.\nonumber \\
\eea
\par
The Euler-Lagrange equations
derived from the energy functional (\ref{eq:O3energy})
are nonlinear \textit{PDE}'s, so in most cases one must resort to 
numerical techniques in order to solve them. In our 
approach, the minimal energy configuration of a baby skyrmion of charge B 
and a given set of values $\mu, \kappa, s$
is found by a full-field relaxation method, which we describe in more detail in the Appendix.

\subsection{Results}
Applying the minimization procedure, 
we obtain the static solutions of the model for $1 \leq B \leq 5$.
Since the parameter space of the model is effectively two dimensional (as discussed earlier),
without loss of generality 
we fix the potential strength at $\mu^2=0.1$ throughout,
and the $s$-$\kappa$ parameter space is scanned in
the region $0<s \leq 4$, $0.01 \leq \kappa^2 \leq 1$.

\subsubsection{Charge-one skyrmions}
\begin{figure}[htp!] \begin{center} 
\includegraphics[angle=0,scale=1,width=1.1\textwidth]{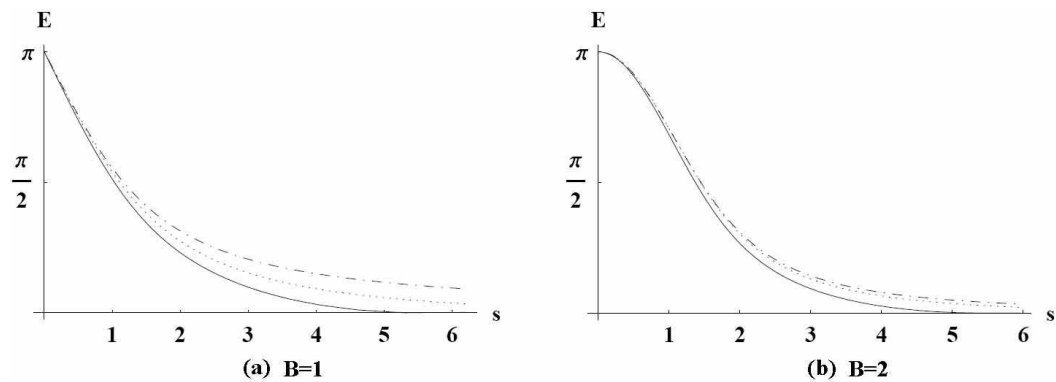}
\caption{Profile functions of the $B=1$ (left) and $B=2$ (right) skyrmions
for $s=0.5$ (solid), $s=1$ (dotted) and $s=2$ (dot-dashed). Here $\kappa$ is fixed at $\kappa^2=0.25$.}
\label{profB1}
\end{center} \end{figure}

\begin{figure}[htp!] \begin{center}
\includegraphics[angle=0,scale=1,width=1\textwidth]{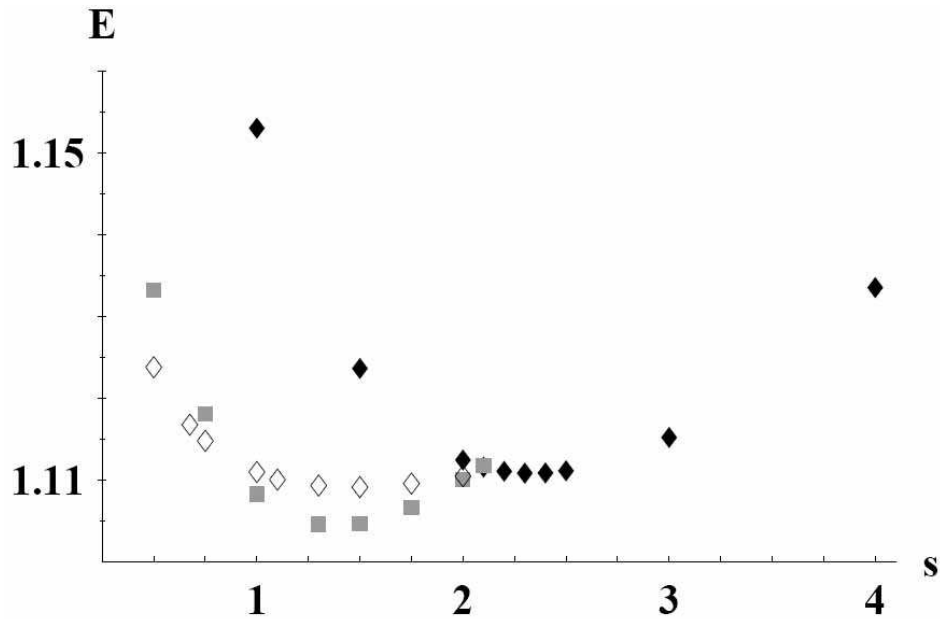}
\caption{Total energies (divided by $4 \pi B$) of the charge-one ({\scriptsize $\blacklozenge$}) charge-two 
({\small \color{Gray}{$\blacksquare$}})
and charge-three ({\small $\lozenge$}) skyrmions 
 as a function of the parameter $s$ for $\kappa^2=0.05$.
Each of the energy graphs attains a minimal value at some $s$. At $s \approx 2$ the 
energy-per-topological-charge of the charge-two and charge-three solutions reaches
the charge-one energy (from below), and stable solutions are no longer observed.}
\label{fig:s123}
\end{center} \end{figure}

In the charge-one sector, the solutions for every value of $s$ and $\kappa$
are stable rotationally-symmetric configurations. 
Figure \ref{profB1}a shows the obtained profile functions of the $B=1$ solution for 
different values of $s$ with $\kappa$ fixed at $\kappa^2=0.25$. 
Interestingly, the skyrmion energy as a function of $s$ is not monotonic,
but acquires a minimum at $s \approx 2.2$, as is shown in Fig. \ref{fig:s123}. 

\subsubsection{Charge-two skyrmions}
Stable solutions also exist in the $B=2$ sector, but only for $s < 2$.
They are rotationally-symmetric and ring-like, corresponding to two charge-one
skyrmions on top of each other. Figure \ref{profB1}b shows the profile functions of
the stable solutions for different values of $s$ and  $\kappa^2=0.25$. 
\par
\begin{figure}  \begin{center} 
\includegraphics[angle=0,scale=1,width=0.9\textwidth]{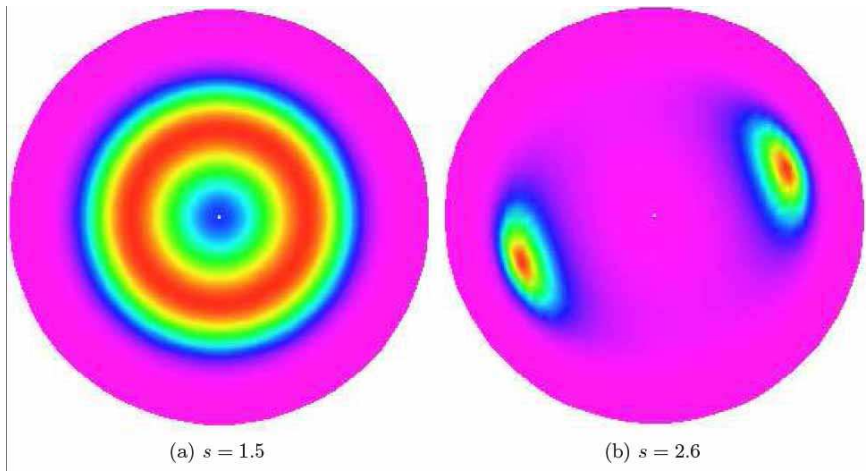}
\caption{Contour plots of the energy distributions (ranging from violet -- low density 
to red -- high density) of the $B=2$ skyrmion for $\kappa^2=1$. 
In the $s<2$ regime, ring-like rotationally-symmetric configurations exist,
corresponding to two charge-one skyrmions on top of each other (left), whereas 
for $s>2$, the charge-two skyrmion splits into two one-charge skyrmions drifting infinitely apart (right).}
\label{contourB2}
\end{center} \end{figure}

As in the $B=1$ case, the energy of the charge-two skyrmion as a function of $s$ is non-monotonic 
and has a minimum around $s = 1.3$. 
As shown in Fig. \ref{fig:s123}, at $s \approx 2$ the energy of the ring-like configuration reaches the value of 
twice the energy of the charge-one skyrmion and stable configurations cease to exist.
At this point, the skyrmion breaks apart into 
its constituent charge-one skyrmions, which in turn
start drifting away from each other, thus breaking the rotational symmetry of the solution. 
Contour plots of the energy distribution of the $B=2$ skyrmion
are shown in Fig. \ref{contourB2} for $\kappa^2=1$ and for two $s$ values.
While for $s=1.5$ a ring-like stable configuration exists (Fig. \ref{contourB2}a),
for $s=2.6$ the skyrmion breaks apart. The latter case is shown in Fig. \ref{contourB2}b which
is a ``snapshot'' taken while the distance between the individual skyrmions kept growing.
\par
These results are in accord
with corresponding results from previously known studies of both the `old' ($s=1$) model
in which ring-like configurations have been observed \cite{Old1,Old2}, and the holomorphic model 
for which no stable solutions have been found \cite{Holo1,Holo2}.

\par
Rotationally-symmetric charge-two locally stable solutions may also be observed
in the large $s$ regime, including the `holomorphic' $s=4$ case, in which case 
the global minimum in this regime corresponds to 
two infinitely separated charge-one skyrmions. 
The total energy of the rotationally symmetric solutions is
larger than twice the energy of a charge-one skyrmion, indicating
that the split skyrmion is
an energetically more favorable configuration.
We discuss this issue in more detail in the section \ref{sec:res}.

\subsubsection{Charge-three and higher-charge skyrmions}
As with the $B=2$ skyrmion, the existence of stable charge-three
skyrmions was also found to be $s$ dependent.
For any tested value of $\kappa$ in the range 
$0.01 \leq \kappa^2 \leq 1$, we have found that above $s \approx 2$, no stable charge-three solutions exist;
in this region the skyrmion breaks apart into individual skyrmions
drifting further and further away from each other. 
\par
\begin{figure}[hbp!] \begin{center}
\includegraphics[angle=0,scale=1,width=0.85\textwidth]{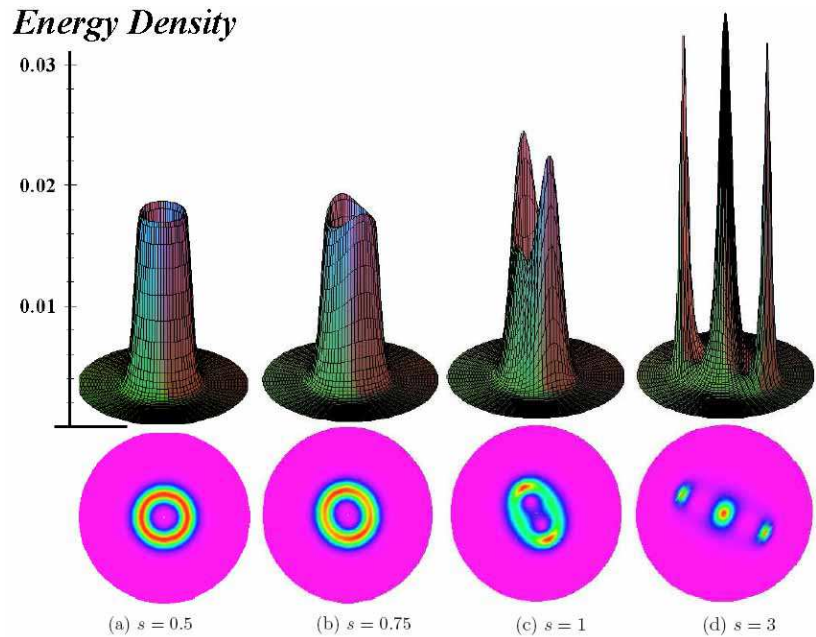}
\caption{Energy densities and corresponding contour plots (ranging from violet -- low density 
to red -- high density) of the $B=3$ skyrmion for fixed $\kappa$ ($\kappa^2=0.01$)
and varying $s$.  
In the $s=0.5$ case, the minimal energy configuration 
is rotationally symmetric, corresponding the three one-skyrmions on top of each other.
For $s=0.75$ and $s=1$ the solutions exhibit only $\mathbb{Z}(2)$ symmetry,
corresponding to partially-overlapping one-skyrmions. For $s=3$ no stable solution exists
and the individual skyrmions are drifting apart. }
\label{fixedKappaB3}
\end{center} \end{figure}

In the $s<2$ region, where stable solutions exist, 
the energy distribution of the charge-three skyrmion
turns out to be highly dependent on both $s$ and $\kappa$.
Keeping $\kappa$ fixed and varying $s$, we find that in the small $s$ regime,
ring-like rotationally-symmetric configurations exist. 
Increasing the value of $s$ results in stable minimal energy configurations
with only $\mathbb{Z}(2)$ symmetry, corresponding to three partially-overlapping charge-one skyrmions in a row,
as shown in Figs \ref{fixedKappaB3}b and \ref{fixedKappaB3}c. 
The energy of the charge-three
skyrmion also has a minimum in $s$, at around $s = 1.5$ (as shown in Fig. \ref{fig:s123}). At $s \approx 2$ 
the energy of the skyrmion becomes larger than three times the
energy of a charge-one skyrmion and stable configurations are no longer obtainable.
This is illustrated in Fig. \ref{fixedKappaB3} which shows 
contour plots of the energy distribution of the $B=3$ skyrmion
for different values of $s$ and fixed $\kappa$. For $s=0.5$ (Fig. \ref{fixedKappaB3}a), the solution
is rotationally symmetric and 
for $s=0.75$ and $s=1$ (Figs \ref{fixedKappaB3}b and \ref{fixedKappaB3}c respectively)
the rotational symmetry 
of the solution is broken and only $\mathbb{Z}(2)$ symmetry remains. At $s=3$, no stable solution exists. 
The latter case is shown in Fig. \ref{fixedKappaB3}d which
is a ``snapshot'' taken while the distance between the individual skyrmions
kept growing.
\par
\begin{figure}[htp!] \begin{center}
\includegraphics[angle=0,scale=1,width=0.99\textwidth]{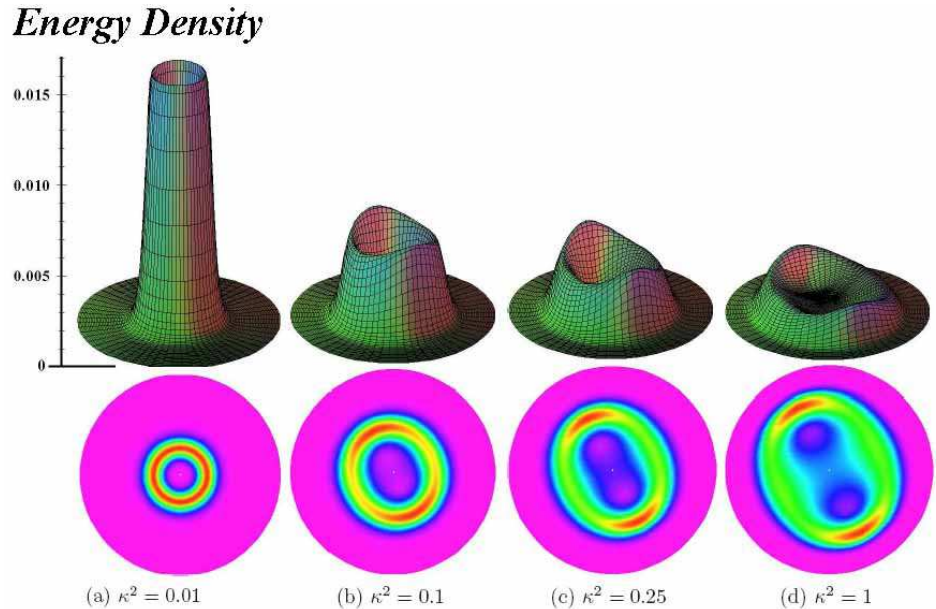}
\caption{Energy densities and corresponding contour plots (ranging from violet -- low density 
to red -- high density) of the $B=3$ skyrmion for fixed $s$ ($s=0.5$)
and varying $\kappa$. At low $\kappa$, the minimal energy configuration is rotationally
symmetric. As $\kappa$ is increased, breaking of rotational symmetry appears,
and only $\mathbb{Z}(2)$ symmetry remains.}
\label{fixedSB3}
\end{center} \end{figure}

The dependence of the skyrmion solutions 
on the value of $\kappa$ with fixed $s$
show the following behavior: While for small $\kappa$ the minimal energy configurations are rotationally-symmetric,
increasing its value results
in an increasingly larger rotational symmetry breaking. This is illustrated in
Fig. \ref{fixedSB3}. 

\begin{figure}[htp!] \begin{center} 
\includegraphics[angle=0,scale=1,width=0.9\textwidth]{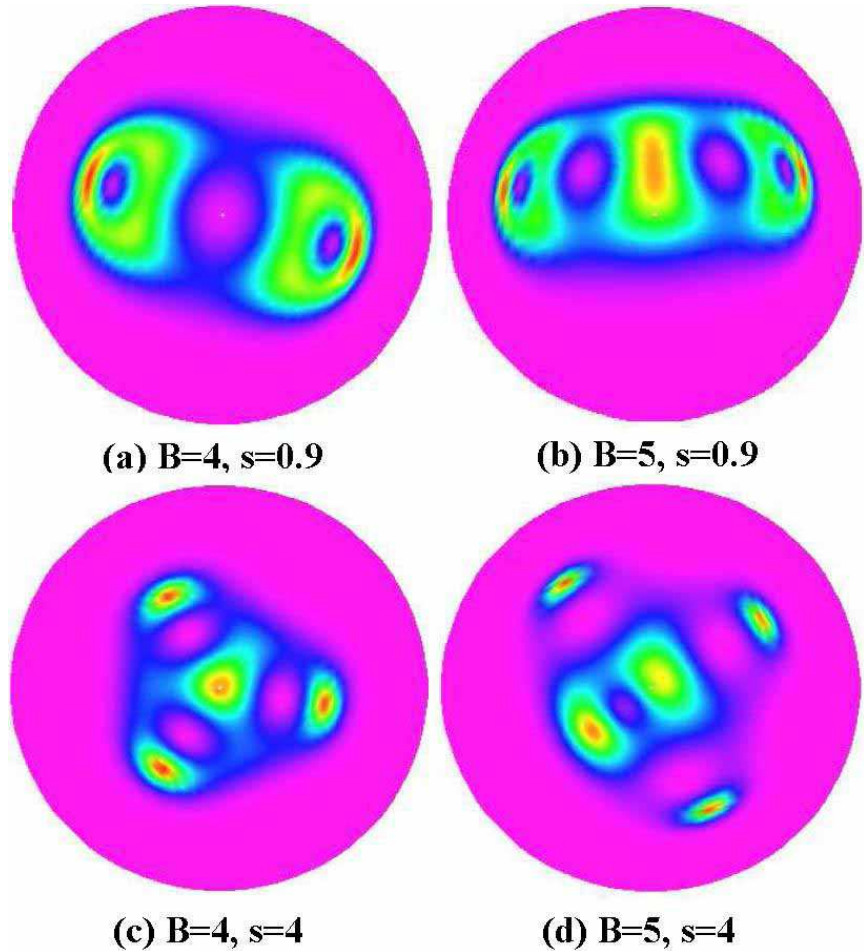}
\caption{Contour plots of the energy distributions (ranging from violet -- low density 
to red -- high density) of the $B=4$ and $B=5$ skyrmions
for $s=0.9$ and $s=4$ ($\kappa^2=0.1$). 
In the lower $s$ region stable solutions exist; the upper figures show a $B=4$ skyrmion 
in a bound state of two charge-two skyrmions (left), and a $B=5$ skyrmion 
in a two-one-two configuration. For values of $s$ higher than $2$, 
the multi-skyrmions split into individual one-skyrmions
constantly drifting apart (lower figures).}
\label{Contour45}
\end{center} \end{figure}

The $B=4$ and $B=5$ skyrmion solutions behave similarly
to the $B=3$ solutions. This is illustrated in Fig. \ref{Contour45},
which shows the stable solutions that have been obtained in the $s=0.9$ case 
and the splitting of these skyrmions
into their constituents in the $s=4$ case.

\section{The Lattice Structure of Baby Skyrmions}
The Skyrme model \cite{Skyrme1} may also be used to describe
systems of a few nucleons, and has also been applied to nuclear and quark matter
 \cite{fewNuc1,fewNuc2,fewNuc3}. One of the most complicated aspects
of the physics of hadrons is the behavior of the phase diagram of hadronic matter at finite
density at low or even zero temperature. 
Particularly, the properties of zero-temperature
skyrmions on a lattice are interesting, 
since the behavior of nuclear matter at high densities is now a focus of considerable 
interest.
Within the standard zero-temperature Skyrme model description, 
a crystal of nucleons turns
into a crystal of half nucleons at finite density 
\cite{Opt3d,maxSym3D,halfSkyrme0, halfSkyrme1,halfSkyrme2}.
\par
To study skyrmion crystals one imposes periodic boundary conditions 
on the Skyrme field and works within a unit cell \cite{TopoSol}.
The first attempted construction of a crystal was by Klebanov \cite{Opt3d}, 
using a simple cubic lattice of skyrmions whose symmetries maximize the 
attraction between nearest neighbors. Other symmetries were proposed which 
lead to crystals with slightly lower, but not minimal energy \cite{maxSym3D,halfSkyrme0}.
It is now understood that it is best to arrange the skyrmions initially as a 
face-centered cubic lattice, with their orientations chosen symmetrically to 
give maximal attraction between all nearest neighbors \cite{halfSkyrme1,halfSkyrme2}. 
\par
The baby Skyrme model too has been
studied under various lattice settings
\cite{2DLattice1,2DLattice2,WC1,WC2,WC3}  
and in fact, it is known that the baby skyrmions also 
split into half-skyrmions when placed inside a rectangular lattice \cite{WC1}.
However, as we shall see, the rectangular periodic boundary conditions 
do not yield the true minimal energy configurations over all possible lattice types \cite{HK2}.
This fact is particularly interesting 
both because of its relevance to
quantum Hall systems in two-dimensions, and also because it may be used to conjecture the
crystalline structure of nucleons in three-dimensions. 
\par
In two dimensions there are five lattice types,
as given by the crystallographic restriction theorem \cite{crt}.
In in all of them the fundamental unit cell is a certain type of a parallelogram.
To find the crystalline structure of the baby skyrmions,
we place them 
inside different parallelograms with periodic boundary conditions
and find the minimal energy configurations over all parallelograms of
fixed area (thus keeping the skyrmion density fixed).
As we show later, there is a certain type of parallelogram, namely the hexagonal,
which yields the minimal energy configuration.
In particular, its energy is lower than the known `square-cell' configurations in which the
skyrmion splits into half-skyrmions. As will be pointed out later,
the hexagonal structure revealed here is not unique to the
present model, but also arises in
other solitonic models, such as Ginzburg-Landau vortices \cite{GL},
quantum Hall systems \cite{QHE,QHF1}, and even in the context of $3$D skyrmions \cite{3Dhexlat}.
The reason for this will also be discussed later.
\par
In what follows we review the setup of
our numerical computations, introducing
a systematic approach for the identification of the minimal energy crystalline structure
of baby skyrmions. In section \ref{sec:res} we present the main results of our study
and in section \ref{sec:semiAna}, a somewhat more analytical 
analysis of the problem is presented. 

\subsection{Baby skyrmions inside a parallelogram}
\begin{figure}[ht!] \begin{center}
\includegraphics[angle=0,scale=1,width=0.9\textwidth]{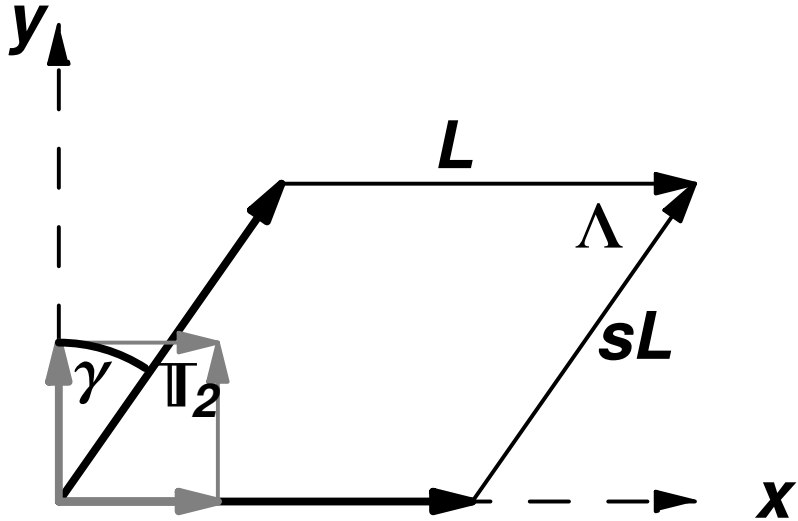}
\caption{\label{par}The parameterization of the fundamental unit-cell parallelogram $\Lambda$ (in black)
and the two-torus $\mathbb{T}_2$ into which it is mapped (in gray).$\hfill{}$}
\end{center} \end{figure}

We find the static solutions of the model by minimizing 
the static energy functional:
\bea
 E =\frac1{2} \int_{\Lambda}  \rmd x \rmd y 
\Big((\partial_{x} \bphi)^2+ (\partial_{y} \bphi)^2 &
+  \kappa^2 (\partial_x \bphi \times \partial_y \bphi)^2
+ 2 \mu^2 (1 -\phi_3) 
\Big)& \,,
\eea
within each topological sector. In this example, we use the `old' model potential term.
In our setup, the integration is over parallelograms, denoted here
by $\Lambda$:   
\bea
\Lambda = \{\alpha_1 (L,0) + \alpha_2 (s L \sin \gamma ,s L \cos \gamma): 0 \leq \alpha_1,\alpha_2<1 \}\,.
\eea
Here $L$ is the length of one side of the parallelogram, $s L$ with $0 < s \leq 1$ is the length of its other side 
and $0 \leq \gamma < \pi/2$ is the angle between  the `$s L$' side and
the vertical to the `$L$' side (as sketched in Fig. \ref{par}). 

Each parallelogram is thus specified by a set $\{L,s,\gamma \}$ and 
the skyrmion density inside a parallelogram is  $\rho = B/(s L^2 \cos \gamma)$,
where $B$ is the topological charge of the skyrmion. 
The periodic boundary conditions are taken into account by 
identifying each of the two opposite sides of a parallelogram as equivalent:
\bea
\bphi(\bx)=\bphi(\bx + n_1 (L,0) +n_2 (s L \sin \gamma ,s L \cos \gamma)) \,,
\eea
with $n_1,n_2 \in \mathbb{Z}$. We are interested in static finite-energy solutions, which in the language of differential
geometry are $\mathbb{T}_2 \mapsto S_2$ maps. These 
are partitioned into homotopy sectors parameterized by
an invariant integral topological charge $B$, the degree of the map, given by:
\bea \label{eq:O3Bnumber}
 B=\frac1{4 \pi} \int_{\Lambda} \rmd x \rmd y 
\left( \bphi \cdot ( \partial_{x} \bphi \times  \partial_{y} \bphi)
\right) \,.
\eea
The static energy $E$ can be shown to satisfy 
\bea \label{ineq}
E \geq 4 \pi B \,,
\eea
with equality possible only in the `pure' $O(3)$ case (i.e., when both 
the Skyrme and the potential terms are absent) \cite{WC1}. 
We note that while in the baby Skyrme model on $\mathbb{R}^2$ with fixed boundary conditions
the potential term is necessary to prevent the solitons from expanding indefinitely,
in our setup it is not required, due to
the periodic boundary conditions \cite{WC1}.
We study the model both with and without the potential term.
\par
The problem in question can be simplified by a linear mapping of
the parallelograms $\Lambda$ into the 
unit-area two-torus $\mathbb{T}_2$.
In the new coordinates, the energy functional becomes
\bea \label{eq:T2energy}
E&=&
\frac{1}{2 s \cos \gamma} 
\int_{\mathbb{T}_2} \rmd x \rmd y 
\left( s^2 (\partial_{x} \bphi)^2
-2 s \sin \gamma  \partial_{x} \bphi \partial_{y} \bphi
+ (\partial_{y} \bphi)^2 \right) \nonumber \\
&+&
\frac{\kappa^2 \rho}{2 B} \int_{\mathbb{T}_2} \rmd x \rmd y 
\left( \partial_x \bphi \times \partial_y \bphi \right)^2
+ \frac{\mu^2 B}{\rho}  \int_{\mathbb{T}_2} \rmd x \rmd y \left(1 -\phi_3 \right)
\,.
\eea
We note that the dependence of the energy on the Skyrme parameters $\kappa$ and $\mu$ and the skyrmion density 
$\rho$ is only through $\kappa^2 \rho$ and $\mu^2/ \rho$. 
\par
In order to find the minimal energy configuration of skyrmions over all
parallelograms with fixed area (equivalently, a specified $\rho$),
we scan the parallelogram parameter space  $\{s,\gamma \}$ and find the parallelogram 
for which the resultant energy is minimal over the parameter space. 
An alternative approach to this problem, which is of a more analytical nature,
may also been implemented. We discuss it in detail in section \ref{sec:semiAna}. 
\subsection{\label{sec:res} Results}
In what follows, we present the minimal energy static skyrmion configurations
over all parallelograms, for various settings:
The `pure' $O(3)$ case, in which both $\kappa$, the Skyrme parameter, and $\mu$, 
the potential coupling, are set to zero, the Skyrme case for which only $\mu=0$, and
the general case for which neither the Skyrme term nor the potential term vanish. 
\par
\begin{figure}[htp!] \begin{center} 
\includegraphics[angle=0,scale=1,width=0.9\textwidth]{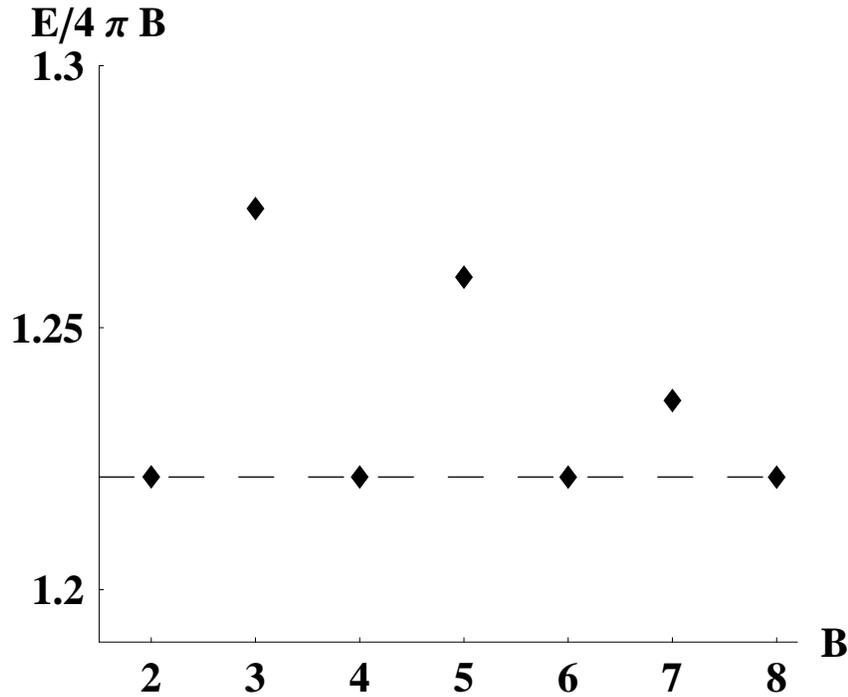}
\caption{\label{difB} Energy per charge of the multi-skyrmion configurations 
as a function of topological charge. The horizontal dashed line was added to guide the eye.
(Here, $\kappa^2=0.03$, $\mu=0$, $\rho=1$, $s=1$ and $\gamma=\pi/6$.)}
\end{center} \end{figure}

In each of these settings, we scanned the parameter space of parallelograms, 
while the skyrmion density $\rho$ was held fixed, 
yielding for each set of $\{ s,\gamma \}$ a minimal energy configuration.
The choice as to how many skyrmions are to be placed inside the unit cells
was made after some preliminary testing in which skyrmions of other charges (up to $B=8$) were also 
examined. The odd-charge minimal-energy configurations turn out to have 
substantially higher energies than even-charge ones,
where among the latter, the charge-two skyrmion is found to be the most fundamental,
as it is observed that the charge-two configuration 
is a `building-block' of the higher-charge configurations.
This is illustrated in Fig. \ref{difB} in which the typical behavior of
the multi-skyrmion energies as a function of 
topological charge is shown.

\subsubsection{The pure $O(3)$ case ($\kappa=\mu=0$)}
\begin{figure}[htp!] \begin{center} 
\includegraphics[angle=0,scale=1,width=0.9\textwidth]{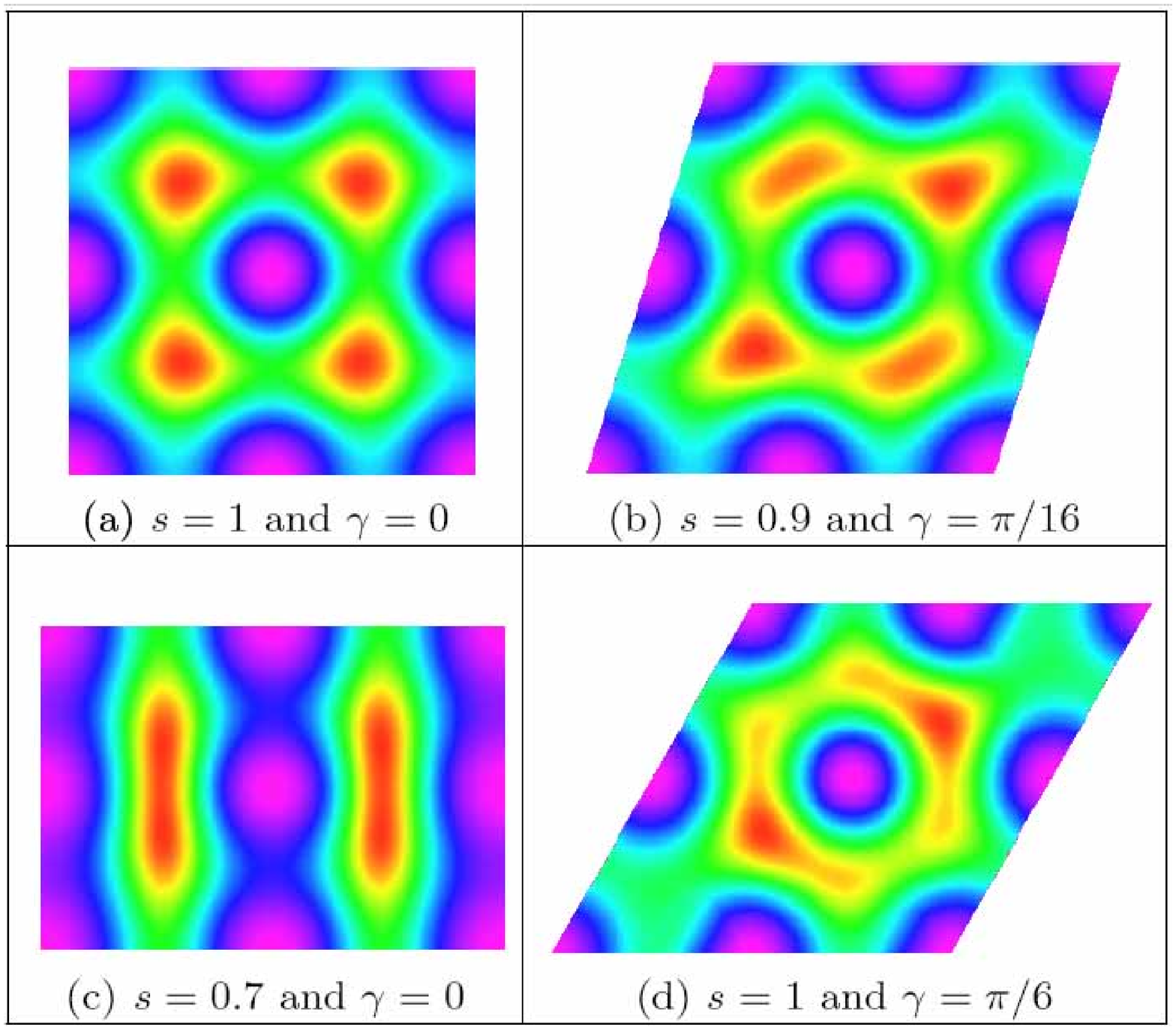}
\caption{\label{cK0}Charge-two skyrmions in the pure $O(3)$ case: 
Contour plots of the charge densities ranging from violet
(low density) to red (high density)
for various parallelogram settings, 
all of which saturate the energy bound $E=4 \pi B=8 \pi$.}
\end{center} \end{figure}

The pure $O(3)$ case corresponds to setting both $\kappa$ and $\mu$
to zero. In this case, analytic solutions in terms of Weierstrass elliptic functions may be 
found \cite{WC1,WC2,WC3} and the minimal energy configurations, 
in all parallelogram settings, saturate the energy bound in (\ref{ineq}) 
giving $E=4 \pi B$. 
Thus, comparison of our numerical results with the analytic solutions serves as a useful check 
on the precision of our numerical procedure.
The agreement is to $6$ significant digits. 
Contour plots of the charge densities for
different parallelogram settings for the charge-two skyrmions are shown in Fig. \ref{cK0},
all of them of equal energy $E/8 \pi =1$.

\subsubsection{The Skyrme case ($\kappa \neq 0, \mu = 0$)}
\begin{figure}[htp!] \begin{center} 
\includegraphics[angle=0,scale=1,width=0.9\textwidth]{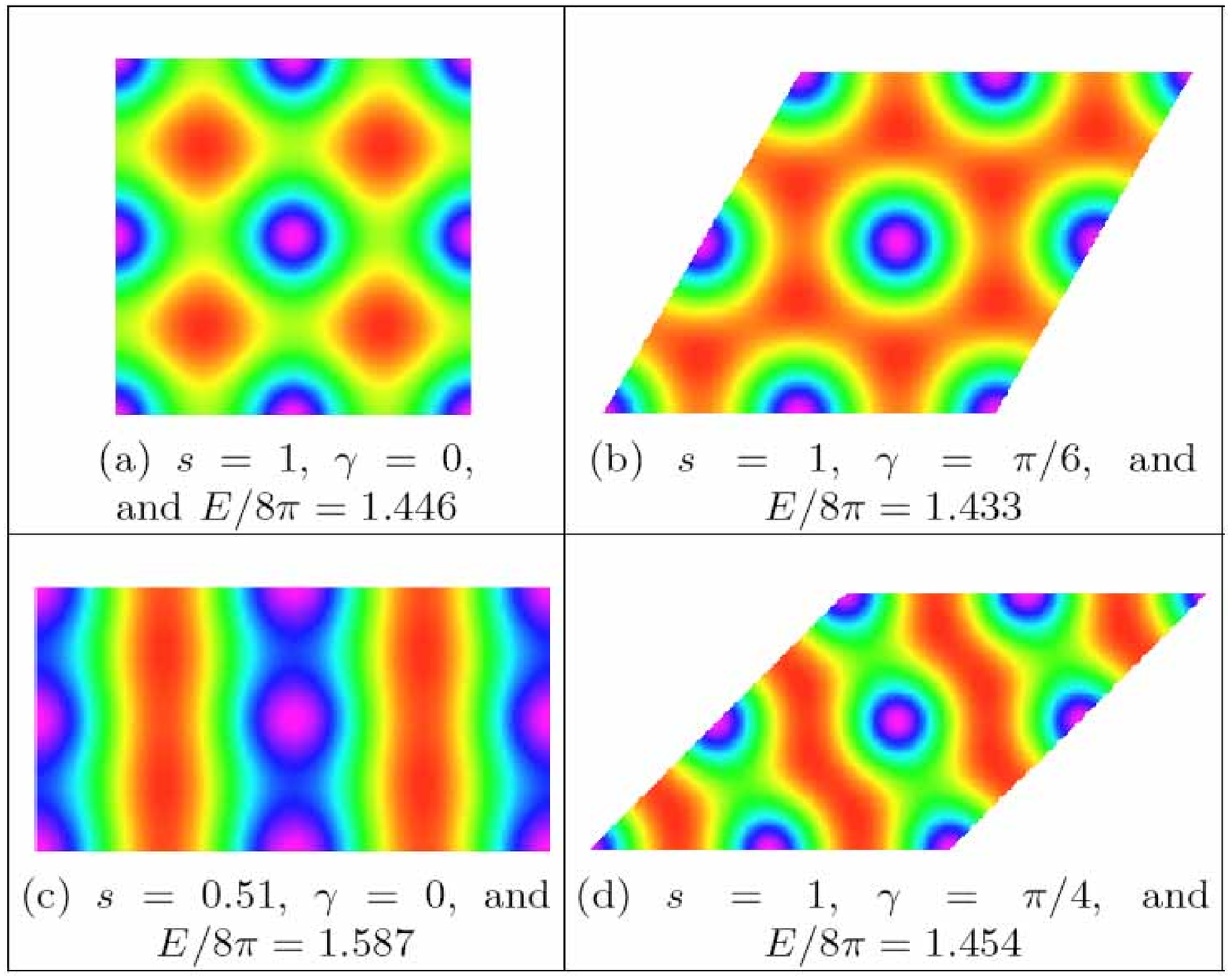}
\caption{\label{contourB2h}Charge-two skyrmions in the Skyrme case with $\kappa^2=0.03$ and $\rho=2$: 
Contour plots of the charge densities
for the hexagonal, square and other fundamental cells 
ranging from violet (low density) to red (high density).
As the captions of the individual subfigures indicate, the hexagonal configuration  
has the lowest energy.$\hfill{}$}
\end{center} \end{figure}

As pointed out earlier, for $\mu=0$ the dependence of the energy functional on
the Skyrme parameter $\kappa$ is only through $\kappa^2 \rho$, 
so without loss of generality we vary $\rho$ and fix $\kappa^2=0.03$ throughout (this
particular choice for $\kappa$ was made for numerical convenience).
Minimization of the
energy functional (\ref{eq:T2energy}) over all parallelograms yields the following.
For any fixed density $\rho$, the minimal energy 
is obtained for $s = 1$ and $\gamma =\pi/6$. This set of values 
corresponds to the `hexagonal' or `equilateral triangular' lattice. 
In this configuration, any three adjacent zero-energy loci (or `holes')
are the vertices of equilateral triangles, and 
eight distinct high-density peaks are observed (as shown in Fig. \ref{contourB2h}b).
This configuration can thus be interpreted as the splitting of the two-skyrmion into 
eight quarter-skyrmions.
This result is independent of the skyrmion density $\rho$. 

In particular, the well-studied square-cell minimal energy configuration (Fig. \ref{contourB2h}a), in which
the two-skyrmion splits into four half-skyrmions,
has higher energy than the hexagonal case. Figure \ref{contourB2h}
shows the total energies (divided by $8 \pi$) and 
the corresponding contour plots of charge densities of the hexagonal, square and other 
configurations (for comparison), all of them with $\rho=2$.

\begin{figure}[htp!] \begin{center} 
\includegraphics[angle=0,scale=1,width=1\textwidth]{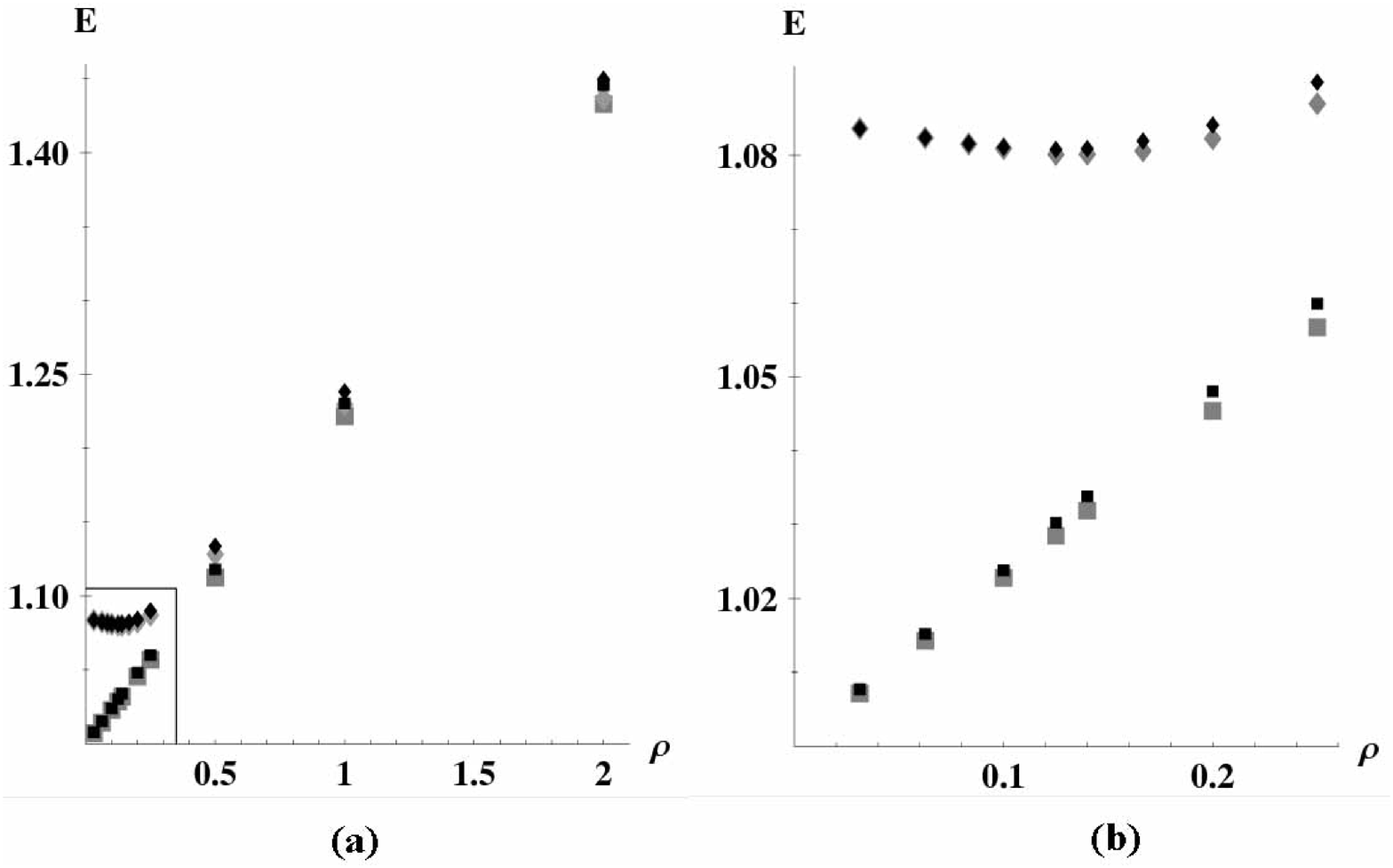}
\caption{\label{Energy} Total energy $E$ (divided by $8 \pi$) of the charge-two skyrmion
in the hexagonal lattice ({\small \color{Gray}{$\blacksquare$}} -- Skyrme case and {\small \color{Gray}{$\blacklozenge$}} -- general case)
and in the square lattice ( {\scriptsize $\blacksquare$} -- Skyrme case and {\scriptsize $\blacklozenge$} -- general case) 
as function of the skyrmion density (in the Skyrme case, $\kappa^2=0.03$ and in the general case 
$\kappa^2=0.03$ and $\mu^2=0.1$).
Note the existence of 
an optimal density (at $\rho \approx 0.14$) in the general case, for which
the energy attains a global minimum.
Figure (b) is an enlargement of the lower left corner of figure (a).
$\hfill{}$}
\end{center} \end{figure}

The total energy of the skyrmions in the hexagonal setting  
turns out to be linearly proportional to the density of the 
skyrmions, reflecting the scale invariance of the model (Fig. \ref{Energy}). 
In particular, the global minimum of $E=4 \pi B =8 \pi$ is reached when $\rho \to 0$.  
This is expected since setting $\rho=0$
is equivalent to setting the Skyrme parameter $\kappa$ to zero, in which case
the model is effectively pure $O(3)$ and inequality (\ref{ineq}) is saturated.

\subsubsection{\label{subsec:general} The general case ($\kappa \neq 0$, $\mu \neq 0$)}
The hexagonal setting turns out to be the 
energetically favorable also in the general case.
Moreover, since in this case the skyrmion has a 
definite size (as is demonstrated by the $\rho$ dependence in the energy functional),
the skyrmion structure is different at low and at high densities and a phase transition occurs at 
a certain critical density.
While at low densities the individual skyrmions are isolated from each other,
at high densities they fuse together, forming the quarter-skyrmion crystal, as in
the Skyrme case reported above.
As the density $\rho$ decreases, or equivalently the value of $\mu$ increases,
the size of the skyrmions becomes small compared to the cell size. 
The exact shape of the lattice loses its effects and the differences in
energy among the various lattice types become very small. This is illustrated in
Fig. \ref{Energy}.
\par
\begin{figure}[htp!] \begin{center} 
\includegraphics[angle=0,scale=1,width=0.9\textwidth]{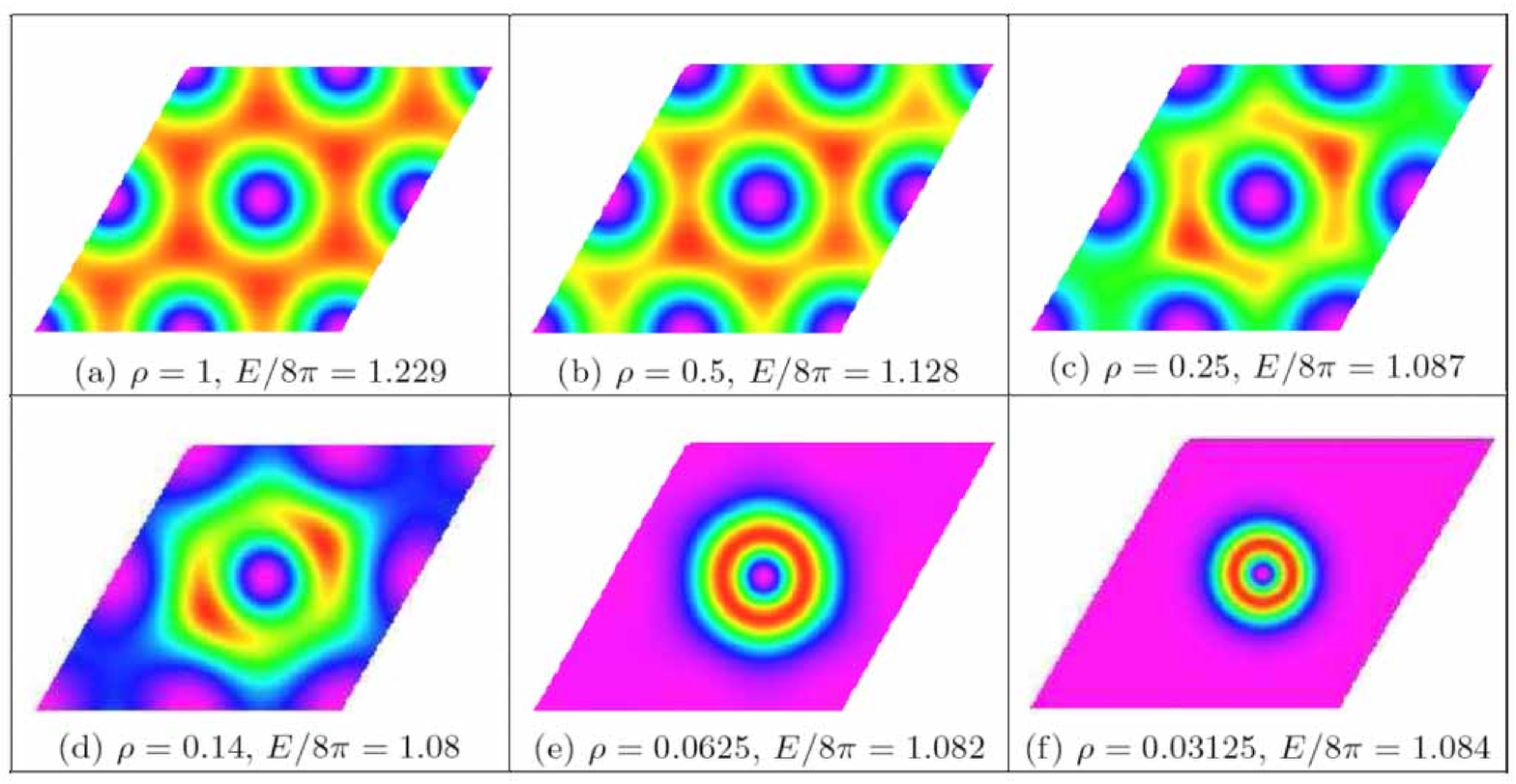}
\caption{\label{contourG2} Charge-two skyrmions in the general case with \hbox{$\mu^2=0.1$} and $\kappa^2=0.03$: 
Contour plots of the charge densities of the minimal-energy configurations
in the hexagonal setting for different densities. Here,
the energetically most favorable density is $\rho= 0.14$.
The plot colors range from violet (low density) to red (high density).$\hfill{}$}
\end{center} \end{figure}

Due to the finite size of the skyrmion, there is
an optimal density for which the energy is minimal among all densities. 
Figure \ref{contourG2} shows the contour plots of the charge density 
of the charge-two skyrmion for several densities 
with $\kappa^2=0.03$ and $\mu^2=0.1$. The energy of the skyrmion is minimal
for $\rho \approx 0.14$ (Fig. \ref{Energy}). 

\subsection{\label{sec:semiAna} Semi-analytical approach}
The energy functional (\ref{eq:T2energy}) depends both on the Skyrme field $\bphi$ and on the parallelogram parameters
$\gamma$ and $s$. Formally, the minimal energy configuration over all parallelograms may be obtained by 
functional differentiation with respect to $\bphi$ and regular
differentiation with respect to $\gamma$ and $s$.
However, since the resulting equations are very complicated, 
a direct numerical solution is quite hard.
Nonetheless, some analytical progress may be achieved in the following way. 
As a first step, we differentiate the energy functional (\ref{eq:T2energy}) only with respect to $\gamma$ and $s$:
\bea \label{eq:SGder}
\frac{\partial E}{\partial \gamma} &=& \frac{1}{2 s \cos^2 \gamma} \left(
\sin \gamma (\mathcal{E}^{yy} + s^2 \mathcal{E}^{xx}) -2 s \mathcal{E}^{xy} \right)  =0 \,, \nonumber\\*
\frac{\partial E}{\partial s}      &=&   \frac{1}{2 s^2 \cos \gamma} (\mathcal{E}^{yy} - s^2 \mathcal{E}^{xx}) =0 \,,
\eea
where $\mathcal{E}^{ij}=\int_{\mathbb{T}_2} \rmd x \rmd y  (\partial_{i} \bphi  \cdot \partial_{j} \bphi)$ and
$i,j \in \{ x,y \}$.
Solving these equations for $\gamma$ and $s$ yields
\bea \label{eq:sGammaMin}
s           &=& \sqrt{\frac{\mathcal{E}^{yy}}{\mathcal{E}^{xx}}} \,, \\\nonumber
\sin \gamma &=& \frac{\mathcal{E}^{xy}}{\sqrt{\mathcal{E}^{xx} \mathcal{E}^{yy}}} \,. 
\eea
Substituting these expressions 
into the energy functional (\ref{eq:T2energy}), we arrive at a `reduced' functional
\bea \label{eq:redE}
E= \sqrt{\mathcal{E}^{xx} \mathcal{E}^{yy} -{(\mathcal{E}^{xy})}^2} +\frac{\kappa^2 \rho}{2 B} \mathcal{E}_{\textrm{sk}} 
+\frac{\mu^2 B}{\rho} \mathcal{E}_{\textrm{pot}}\,,
\eea
where 
$\mathcal{E}_{\textrm{sk}}=\int_{\mathbb{T}_2} \rmd x \rmd y  \left( \partial_x \bphi \times \partial_y \bphi \right)^2$
is the Skyrme energy and $\mathcal{E}_{\textrm{pot}}=\int_{\mathbb{T}_2} \rmd x \rmd y  \left( 1-\phi_3 \right)$
is the potential energy.
Now that both $\gamma$ and $s$ are eliminated from the resultant expression, 
and the conditions for their optimization are built into the functional,
the numerical minimization is carried out.
We note here, however, that the procedure presented above should be treated with caution.
This is since Eqs. (\ref{eq:sGammaMin}) are only extremum conditions, and may correspond
to a maximum or saddle-point. 
Hence, it is important to confirm any results obtained using this method
by comparing them with corresponding results obtained from the method described in the previous section. 
\par
It is therefore reassuring that numerical minimization of the reduced functional (\ref{eq:redE})
gives $\sin \gamma = 0.498$ ($\gamma \approx \pi/6$) and $s=1$ (both for the Skyrme case and the general case), 
confirming the results presented in the previous section.
\par
In the general ($\mu \neq 0$) case, the energy functional (\ref{eq:redE}) may be further differentiated
with respect to the skyrmion density
$\rho$ to obtain the optimal density for which the skyrmion energy is minimal.
Differentiating with respect to $\rho$, and substituting the obtained 
expression into the energy functional, results in the functional 
\bea \label{eq:redE2}
E= \sqrt{\mathcal{E}^{xx} \mathcal{E}^{yy} -{(\mathcal{E}^{xy})}^2} 
+\kappa \mu \sqrt{ 2 \mathcal{E}_{\textrm{sk}} \mathcal{E}_{\textrm{pot}}}  \,.
\eea
Numerical minimization of the above expression for $\kappa^2=0.03$ 
and various $\mu$ values ($0.1 \leq \mu^2 \leq 10$)
yields the hexagonal setting as in the Skyrme case.
In particular, for $\mu^2=0.1$ the optimal density 
turns out to be $\rho \approx 0.14$, in accord
with results presented in Sec. \ref{subsec:general}.

\subsection{\label{sec:summary} Further remarks}
\par
As pointed out earlier,
the special role of the hexagonal lattice revealed here is not 
unique to the baby Skyrme model, but in fact arises 
in other solitonic models.
In the context of Skyrme models, the existence of a hexagonal two dimensional structure of $3$D skyrmions
has also been found by Battye and Sutcliffe \cite{3Dhexlat}, where it has already been noted that 
energetically, the optimum infinite planar structure of $3$D skyrmions 
is the hexagonal lattice, which resembles the 
structure of a graphite sheet, the most stable form of carbon thermodynamically \cite{TopoSol}.  
Other examples in which the hexagonal structure is revealed are Ginzburg-Landau vortices which 
are known to have lower energy 
in a hexagonal configuration than on a square lattice \cite{GL}.
Thus, it should not come as a surprise that the hexagonal structure 
is found to be the most favorable in the baby Skyrme model. 
\par
As briefly noted in the opening paragraphs of this section,
a certain type of baby skyrmions also arise in quantum Hall systems
as low-energy excitations of the ground state near ferromagnetic filling factors (notably $1$ and $1/3$) 
\cite{QHE}. It has been pointed out that this state contains a finite density of skyrmions \cite{Brey},
and in fact the hexagonal configuration has been suggested as a candidate for their lattice structure \cite{QHF1}. 
Our results may therefore serve as a supporting evidence in that direction, 
although a more detailed analysis is in order.
\par
Our results also raise some interesting questions.
The first is how the dynamical properties of baby skyrmions on the
hexagonal lattice differ from their behavior in the usual rectangular lattice.
Another question has to do with their behavior in non-zero-temperature.
\par 
One may also wonder whether and how these results can be generalized to 
 the $3$D case.
Is the face-centered cubic lattice indeed the minimal energy crystalline structure of $3$D skyrmions
among all parallelepiped lattices? If not, what is the minimal energy structure,
and how do these results depend on the presence of a mass term? 
These questions await a systematic study.

\section{Baby Skyrmions on the Two-Sphere}
Although skyrmions were originally introduced to describe baryons in three 
spatial dimensions \cite{Skyrme1},
they have been shown to exist for a very wide 
class of geometries \cite{Geom}, specifically cylinders, 
two-spheres and three-spheres \cite{cyl,2Sphere1,2Sphere2,3Sphere1,3Sphere2}.
\par
Here, we consider a baby Skyrme model on the two-sphere
\footnote{This type of model has been studied before \cite{2Sphere1,2Sphere2},
although only rotationally-symmetric configurations have been considered.}.
We compute the 
full-field minimal energy solutions of the
model up to charge $14$
and show that they exhibit complex multi-skyrmion solutions
closely related to the skyrmion solutions of the $3$D model with the same
topological charge. 
To obtain the minimum energy configurations, we apply two completely different methods. 
One is the full-field relaxation method, with which exact numerical solutions of the model
are obtained, and the other is a rational map approximation scheme,
which as we show yields very good approximate solutions.
\par
In an exact analogy to the $3$D Skyrme model,
the charge-one skyrmion has a spherical energy distribution, 
the charge-two skyrmion is  
toroidal, and skyrmions with higher charge all have point
symmetries which are subgroups of O(3). 
As we shall see, it is not a coincidence that the symmetries of these solutions are the same as those of
the $3$D skyrmions.

\subsection{The baby Skyrme model on the two-sphere}
The model in question is a baby Skyrme model 
in which both the domain and target are two-spheres.
The Lagrangian density here is simply
\bea \label{O3lagrangian}
\mathcal{L}= \frac1{2} \partial_{\mu} \bphi \cdot \partial^{\mu} \bphi
+ \frac{\kappa^2}{2} \big[ 
(\partial_{\mu} \bphi \cdot \partial^{\mu} \bphi)^2-
(\partial_{\mu} \bphi \cdot \partial_{\nu} \bphi)
(\partial^{\mu} \bphi \cdot \partial^{\nu} \bphi)
\big]\,,
\eea
with metric $\rmd s^2=d t^2 - \rmd \theta^2 -\sin^2 \theta \, \rmd \varphi^2$,
where $\theta$ is the polar angle $\in [0,\pi]$ and $\varphi$
is the azimuthal angle $\in [0,2 \pi)$.
The Lagrangian of this model
is invariant under rotations in both domain and the target spaces,
possessing an 
$\,O(3)_{\textrm{{\small domain}}} \times O(3)_{\textrm{{\small target}}}\,$ symmetry.
As noted earlier, in flat two-dimensional space an additional potential term is necessary to
ensure the existence of stable finite-size solutions. Without it, the repulsive
effect of the Skyrme term causes the skyrmions to expand indefinitely.
In the present model, however, the finite geometry of the sphere acts as a stabilizer,
so a potential term is not required. 
Furthermore, stable solutions exist even without the Skyrme term.
In the latter case, we obtain the well known $O(3)$ (or $\mathbb{CP}^1$) nonlinear sigma model \cite{O3a}.\par
As before, the field $\bphi$ in this model is an $S^2 \to S^2$ mapping, so the relevant homotopy group 
is $\pi_2(S^2)=\mathbb{Z}$, implying that each field configuration is 
characterized by an integer topological charge $B$, the topological degree of
the map $\bphi$. In spherical coordinates $B$ is given by
\begin{equation}
B= \frac1{4 \pi } \int \rmd \, \Omega 
	\frac{ \bphi \cdot (\partial_{\theta} \bphi \times \partial_{\varphi} \bphi )}{
	\sin \theta} \,,
\end{equation}
where $\rmd \Omega= \sin \theta \, \rmd \theta \, \rmd \varphi$. 
\par
Static solutions are obtained by minimizing the energy functional
\bea \label{eq:O3energyS2}
E=\frac1{2} \int \rmd \Omega \Big((\partial_{\theta} \bphi)^2 
+ \frac{1}{\sin^2 \theta}( \partial_{\varphi} \bphi)^2 \Big)
+\frac{\kappa^2}{2} \int \rmd \Omega \Big( \frac{(\partial_{\theta} \bphi \times \partial_{\varphi} \bphi)^2}{\sin^2 \theta}
\Big) \,,
\eea
within each topological sector. 
Before proceeding, it is worthwhile to note that setting $\kappa=0$ 
in  Eq. (\ref{eq:O3energyS2}) yields the energy functional
of the $O(3)$ nonlinear sigma model.
The latter has
analytic minimal energy solutions within every topological sector, given by
\begin{equation} \label{eq:rotSymForm}
\bphi=(\sin f(\theta) \cos(B \varphi),\sin f(\theta) \sin(B \varphi),\cos f(\theta))\,,
\end{equation}
where $f(\theta)=\cos^{-1}(1-2(1+(\lambda \tan \theta/2)^{2 B})^{-1})$ with $\lambda$ being some 
positive number \cite{O3a}.
These solutions are not unique, as other solutions with the same energy 
may be obtained by rotating (\ref{eq:rotSymForm}) either in the target or in the domain spaces. 
The energy distributions of these solutions in each sector have total
energy $E_{B}=4 \pi B$. \par
Analytic solutions also exist for the energy 
functional (\ref{eq:O3energyS2}) with the Skyrme term only. 
They too have the rotationally symmetric form (\ref{eq:rotSymForm})
with $f(\theta)=\theta$ and total energy $E_{B}=4 \pi B^2$.
They can be shown to be the global minima
by the following Cauchy-Schwartz inequality:
\bea
\left( \frac1{4 \pi} \int \rmd \Omega \frac{ \bphi \cdot (\partial_{\theta} \bphi \times \partial_{\varphi} \bphi )}{
	\sin \theta} \right)^2 
	\leq \left( 
	\frac1{4 \pi} \int \rmd \Omega \bphi^2
	\right) \cdot 	\left( \frac1{4 \pi} \int \rmd \Omega
	(\frac{\partial_{\theta} \bphi \times \partial_{\varphi} \bphi}{
	\sin \theta})^2
	\right) \,.\nonumber \\
\eea
The left-hand side is simply $B^2$ and the first term in parenthesis 
on the right-hand side integrates to 1. 
Noting that the second term in the RHS is the Skyrme energy (without the $\kappa^2/2$ factor), 
the inequality
reads $E \geq 4 \pi  B^2$,
with an equality for the rotationally-symmetric solutions.

\subsection{\label{sec:statSol} Static solutions}
In general, if both the kinetic and Skyrme terms are present,
static solutions of the model cannot be obtained analytically. 
This is with the exception of the $B=1$ skyrmion which has
an analytic ``hedgehog'' solution
\begin{equation}
\bphi_{[B=1]}=(\sin \theta \cos \varphi, \sin \theta \sin \varphi, \cos \theta)\,,
\end{equation}  
with total energy 
$\displaystyle \frac{E}{4 \pi}=1+\frac{\kappa^2}{2}\deepstrut$. 
\par
For skyrmions with higher charge, we find the minimal energy configurations
by utilizing the full-field relaxation method described earlier.
In parallel, we also apply the rational map approximation method, originally developed
for the $3$D Skyrme model and compare the results with the relaxation method.
Let us briefly discuss the rational map approximation method: 
computing the minimum energy configurations 
using  the full nonlinear energy functional is 
a procedure which is both time-consuming and resource-hungry.
To circumvent these problems, the rational map ansatz scheme has been devised.
First introduced by Houghton, Manton and Sutcliffe \cite{Rmaps3DSk1}, this scheme
has been used in obtaining approximate solutions
to the $3$D Skyrme model using rational maps between Riemann spheres. 
Although this representation is not exact, it drastically 
reduces the number of degrees of freedom in the
problem, allowing computations to
take place in a relatively short amount of time. 
The results in the case of $3$D Skyrme model are known
to be quite accurate.
\par
Application of the approximation, begins with expressing
points on the base sphere by the Riemann sphere coordinate
$\displaystyle z=\tan \frac{\theta}{2} \rme^{i \varphi}$. 
The complex-valued  function $R(z)$ is a rational map of
degree $B$ between Riemann spheres
\begin{equation}
R(z)=\frac{p(z)}{q(z)}\,,
\end{equation}
where $p(z)$ and $q(z)$ are polynomials in $z$, such that
$\max[\mbox{deg}(p),\mbox{deg}(q)]=B$,  and $p$ and $q$ have no common factors.  
Given such a rational map, the ansatz for the field triplet is
\begin{equation} \label{eq:Rmap}
\bphi=( \frac{R+ \bar{R}}{1+|R|^2}, i \frac{R- \bar{R}}{1+|R|^2}, \frac{1-|R|^2}{1+|R|^2})\,.
\end{equation}
It can be shown that rational maps of degree $B$ correspond to field configurations 
with charge $B$  \cite{Rmaps3DSk1}. Substitution of the ansatz (\ref{eq:Rmap}) into the 
energy functional (\ref{eq:O3energyS2}) results in the simple expression
\begin{equation} \label{eq:Ermap}
\frac1{4 \pi} E=B  + \frac{\kappa^2}{2} \mathcal{I} \,,
\end{equation}
with
\begin{equation} \label{eq:Iint}
\mathcal{I}=\frac{1}{4\pi}\int \bigg(
\frac{1+\vert z\vert^2}{1+\vert R\vert^2}
\bigg \vert\frac{d R}{d z}\bigg\vert\bigg)^4 \frac{2i \ \rmd z  \rmd \bar z
}{(1+\vert z\vert^2)^2}\,.
\end{equation}
Note that in the $\kappa \to 0$ limit, where our model reduces to the $O(3)$ nonlinear sigma model,
the rational maps become exact solutions and 
the minimal energy value $E = 4 \pi B$ is attained.
Furthermore, the minimal energy is reached 
independently of the specific details of the map (apart from its degree), i.e.,  
all rational maps of a given degree are minimal energy configurations in the topological sector
corresponding to this degree. This is a reflection of the scale- and the rotational invariance
of the $O(3)$ model. 
\par
In the general case where $\kappa \neq 0$, the situation is different.
Here, minimizing the energy (\ref{eq:Ermap}) 
requires finding the rational map which minimizes the functional
$\mathcal{I}$. As we discuss in the next section,
the expression for $\mathcal{I}$ given in Eq. (\ref{eq:Iint})
is encountered in the application of the rational map in
the context of $3$D skyrmions, where the procedure of 
minimizing $\mathcal{I}$ over all rational maps of the various degrees
has been used \cite{Rmaps3DSk1,Rmaps3DSk2,Rmaps3DSk3}. 
Here we redo the calculations, utilizing a relaxation method:
to obtain the rational map of degree $B$ that minimizes 
$\mathcal{I}$, 
we start off with a rational map of degree $B$, with the real and imaginary parts 
of the coefficients of $p(z)$ and $q(z)$ assigned random values from  the segment $[-1,1]$. 
Solutions are obtained 
by relaxing the map until a minimum of $\mathcal{I}$ is reached.

\subsection{Relation to the 3D Skyrme model}
In the $3$D Skyrme model,
the rational map ansatz can be thought of
as taken in two steps. 
First, the radial coordinate is 
separated from the angular coordinates by
taking the SU(2) Skyrme field $U(r,\theta,\varphi)$ to be of the form
\begin{equation} \label{ansatz}
U(r,\theta, \varphi)=\exp(if(r) \ \bphi(\theta,\varphi)\cdot \bsigma) \,,
\end{equation}
where $\mbox{\boldmath $\sigma$}=(\sigma_1,\sigma_2,\sigma_3)$ are Pauli matrices,
$f(r)$ is the radial profile function subject to
the boundary conditions  $f(0)=\pi$ and $f(\infty)=0$,
and $\bphi(\theta,\varphi): S^2 \mapsto S^2$ is a normalized vector 
which carries the angular dependence of the field.
In terms of the ansatz (\ref{ansatz}),
the energy of the Skyrme field is

\bea \label{eq:energy3}
E&=&\int 4 \pi {f'}^2  r^2 \rmd r 
+\int 2({ f ' }^2 + 1)\sin^2 f \rmd r 
\int \Big( (\partial_{\theta} \bphi)^2 
+ \frac{1}{\sin^2 \theta}( \partial_{\varphi} \bphi)^2 \Big) \rmd \Omega
\nonumber \\
&+& \int \frac{\sin^4f}{r^2} \rmd r 
 \int \frac{(\partial_{\theta} \bphi \times \partial_{\varphi} \bphi)^2}{\sin^2 \theta} \rmd \Omega.
\eea

Note that the energy functional (\ref{eq:energy3})
is actually the energy functional of our model (\ref{eq:O3energyS2}) once the radial coordinate
is integrated out. Thus, our $2$D model can be thought of as 
a $3$D Skyrme model with a `frozen' radial coordinate.
\par
The essence of the rational map approximation is 
the assumption that $\bphi(\theta,\varphi)$ 
takes the rational map form (\ref{eq:Rmap}),
which in turn leads to a simple expression for the energy
\begin{equation} \label{eq:energy4}
E=4\pi\int \bigg(
r^2 {f'}^2+2B({f'}^2+1)\sin^2 f+\mathcal{I}\frac{\sin^4 f}{r^2}\bigg) \ \rmd r \,, 
\end{equation}
where $\mathcal{I}$ is given in Eq. (\ref{eq:Iint}).
As in the baby model on the two-sphere, minimizing the energy functional 
requires minimizing $\mathcal{I}$ over all maps of degree $B$,
which is then followed by finding the profile function $f(r)$.
\par
Since the symmetries of the $3$D skyrmions are determined solely by the angular dependence 
of the Skyrme field, it should not be too surprising that the solutions of the model discussed here
share the symmetries of the corresponding solutions of the $3$D Skyrme model.

\subsection{Results}
\begin{figure} \begin{center}
\includegraphics[angle=0,scale=1,width=.8\textwidth]{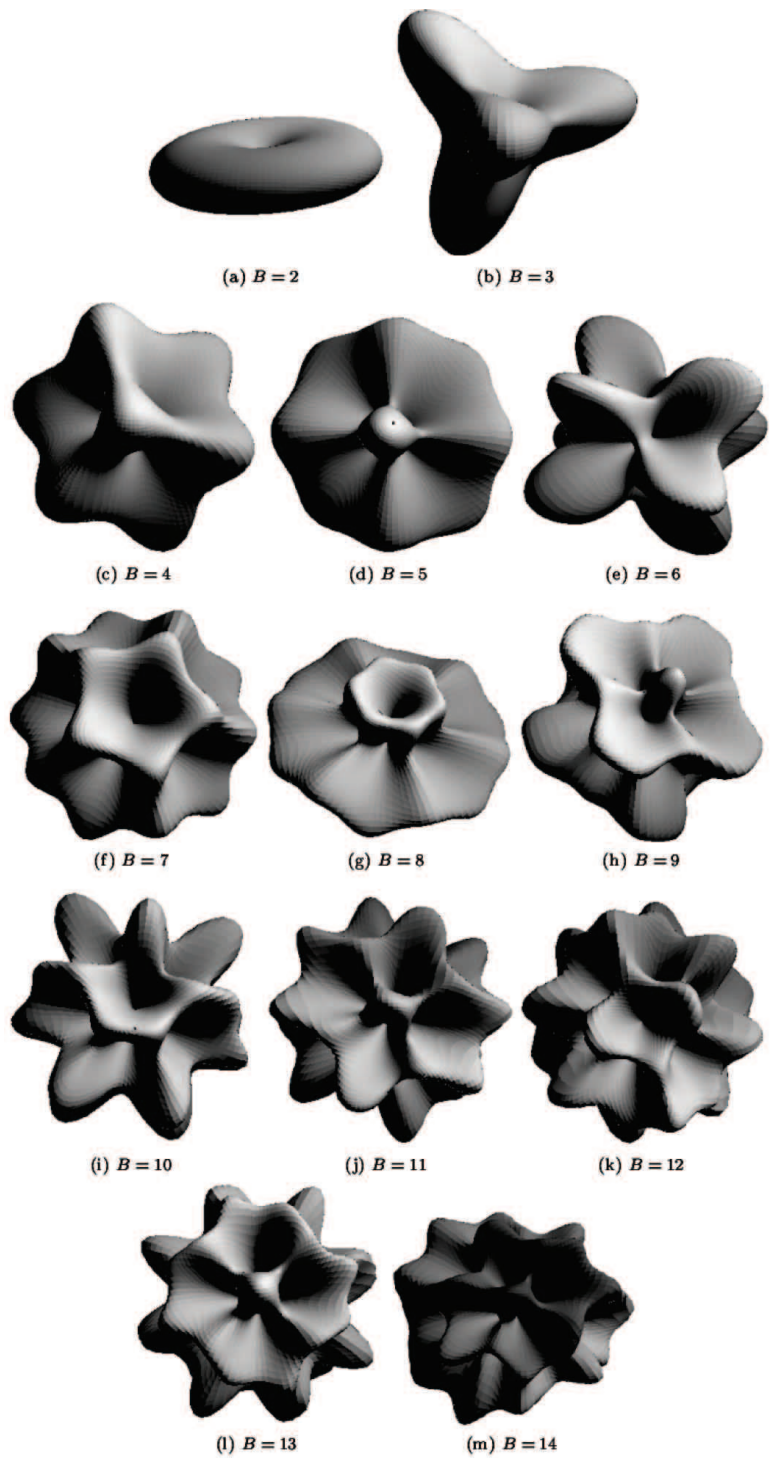}
\caption{\label{Figure1}The energy distributions of the multi-skyrmion solutions for charges \hbox{$2 \leq B \leq 14$} \hbox{($\kappa^2=0.05$).}}
\end{center}
\end{figure}

The configurations obtained from the full-field relaxation method 
are found to have the same symmetries as 
corresponding multi-skyrmions of the $3$D model with the same charge.
The $B=2$ solution is axially
symmetric, whereas higher-charge solutions were all found to have point
symmetries which are subgroups of O(3). For $B=3$ and $B=12$,
the skyrmions have a tetrahedral symmetry. 
The $B=4$ and $B=13$ skyrmions have a cubic symmetry, 
and the $B=7$ is dodecahedral. 
The other skyrmion solutions 
have dihedral symmetries. For $B=5$ and $B=14$ a $D_{2d}$ symmetry, 
for $B=6,9$ and $10$ a $D_{4d}$ symmetry, 
for $B=8$ a $D_{6d}$ symmetry and for $B=11$ a $D_{3h}$ symmetry.
In Fig. ~\ref{Figure1} we show the energy distributions of the obtained solutions for $\kappa^2=0.05$.
\par
While for solutions with $B<8$ the energy density and the charge
density are distributed in distinct peaks,
for solutions with higher charge they are spread in a much more complicated manner.  
The total energies of the solutions (divided by $4 \pi B$) are listed in Table ~\ref{tab1}, 
along with the symmetries of the solutions (again with $\kappa^2=0.05$).
\par
Application of the rational map ansatz yields results
with only slightly higher energies, only about $0.3 \%$ to $3 \%$
above the full-field results.
The calculated values of $\mathcal{I}$ are in agreement with results 
obtained in the context of $3$D skyrmions  \cite{Rmaps3DSk2}. For $9 \leq B \leq 14$,
the rational map approximation yielded slightly less symmetric
solutions than the full-field ones.
Considering the relatively small number of degrees of freedom,
this method all-in-all yields very good approximations. 
The total energies
of the solutions obtained with the rational map approximation 
is also listed in Table ~\ref{tab1}.

\begin{table}[ht]
\tbl{\label{tab1}Total energies (divided by $4 \pi B$)
of the multi-solitons of the model for $\kappa^2=0.05$.}
{\begin{tabular}{@{}llllll}
\toprule
Charge & Total energy & Total energy  & Difference & Symmetry of \\ 
$B$ & Full-field  & Rational maps & in $\%$    & the solution  \\ 
\colrule
$2$  & $1.071$  & $1.073$  	& $0.177$      & Toroidal \\
$3$  & $1.105$  & $1.113$ 	& $0.750$      & Tetrahedral \\
$4$  & $1.125$  & $1.129$ 	& $0.359$ 	& Cubic \\
$5$  & $1.168$  & $1.179$ 	& $0.958$	& $D_{2d}$ \\
$6$  & $1.194$  & $1.211$ 	& $1.426$	& $D_{4d}$ \\
$7$  & $1.209$  & $1.217$ 	& $0.649$	& Icosahedral \\
$8$  & $1.250$  & $1.268$ 	& $1.406$	& $D_{6d}$ \\
$9$  & $1.281$  & $1.304$  	& $1.771$	& $D_{4d}$ \\
$10$ & $1.306$  & $1.332$ 	& $1.991$ 	& $D_{4d}$ \\
$11$ & $1.337$  & $1.366$ 	& $2.224$	& $D_{3h}$ \\
$12$ & $1.360$  & $1.388$ 	& $2.072$	& Tetrahedral \\
$13$ & $1.386$  & $1.415$ 	& $2.137$	& Cubic \\
$14$ & $1.421$  & $1.459$ 	& $2.712$	& $D_2$ \\
\botrule
\end{tabular}}
\end{table}

\newpage
\subsection{Further remarks}
As we have just seen, the baby Skyrme model on the two-sphere 
shares very significant similarities with the $3$D model, especially in terms of multi-skyrmion symmetries.
The fact that the model discussed here is two-dimensional makes it 
simpler to study and perform computations with, when compared 
with the $3$D Skyrme model. 
\par
Some of the results presented above may, at least to some extent,
also be linked to the baby skyrmions which appear in two-dimensional
electron gas systems, exhibiting the quantum Hall effect. 
As briefly noted in the Introduction, baby skyrmions arise in quantum Hall systems
as low-energy excitations of the ground state, near ferromagnetic filling factors (notably $1$ and $1/3$) 
\cite{QHE,QHE2}. 
There, the skyrmion is a twisted two-dimensional configuration of spin, and its topological charge 
corresponds to the number of time the spin rotates by $2 \pi$. 
While for the electron gas, the stability of the soliton 
arises from a balance between the electron-electron Coulomb energy and the Zeeman energy,
in our model the repulsive Skyrme-term energy is balanced 
by the underlying geometry (i.e., the sphere).
The connection between these two models
suggests possible existence of very structured spin textures in quantum Hall systems,
although a more detailed analysis of this analogy is in order. 

\section{Rotating Baby Skyrmions}
We now turn to analyze the phenomenon of spontaneous breaking of rotational symmetry (SBRS)
as it appears in rotating baby skyrmions.
In general, SBRS refers to cases where physical systems which rotate fast enough
deform in a manner which breaks their rotational symmetry, 
a symmetry they posses when static or rotating slowly. 
The recognition that rotating physical systems 
can yield solutions with less symmetry than the governing equations 
is not new. One famous example which dates back to $1834$ is that of
the equilibrium configurations of a rotating fluid mass.  It was  
Jacobi who was first to discover that
if rotated fast enough, a self-gravitating fluid mass can have equilibrium
configurations lacking rotational symmetry.
In modern terminology, Jacobi's asymmetric equilibria appear through a symmetry breaking 
bifurcation from a family of symmetric equilibria as the angular momentum 
of the system increases above a critical value (a ``bifurcation point'') \cite{Lyttle,Chandra}.
Above this critical value, rotationally-symmetric configurations 
are no longer stable, and configurations with a broken rotational symmetry become
energetically favorable.
\par
By now it is widely recognized that symmetry-breaking bifurcations 
in rotating systems are of frequent occurrence and that this is in fact a very general phenomenon,
appearing in a variety of physical settings, among which are
fluid dynamics, star formation, heavy nuclei, chemical reactions, plasmas, 
and biological systems, to mention some diverse examples. 
\par
Recently, SBRS has also been observed in self-gravitating
$N$-body systems \cite{Votyakov1,Votyakov2}, where 
the equilibrium configurations of an $N$-body self-gravitating system enclosed in a finite 3 dimensional spherical volume
have been investigated using a mean-field approach. 
It was shown that when the ratio of the angular momentum of the system to its energy is high, 
spontaneous breaking of rotational symmetry occurs, manifesting itself in the formation of double-cluster
structures. These results have also been confirmed with direct numerical simulations \cite{LMS}.
\par
It is well-known that a large number of phenomena exhibited by many-body systems have
their counterparts and parallels in field theory, which in some sense is
a limiting case of $N$-body systems in the limit $N \to \infty$. 
Since the closest analogues of a lump of matter in field theories are solitons,
the presence of SBRS in self-gravitating $N$-body systems has led us to expect that it 
may also be present in solitonic field theories.
\par
Our main motivation towards studying SBRS in solitons
is that in hadronic physics Skyrme-type solitons 
often provide a fairly
good qualitative description of nucleon properties (See, {\it e.g.}, \cite{revSk1,revSk2}). 
In particular, it is interesting to ask what
happens when such solitons rotate quickly,
because this might shed some light on the non-spherical deformation of 
excited nucleons with high orbital angular momentum, a subject which is now
of considerable interest. 

We shall see that the baby Skyrme model on the two-sphere 
indeed exhibits SBRS, and we will try to understand why this is so \cite{HK4}.
First, we give a brief account for 
the occurrence of SBRS in physical systems in general,
and then use the insights 
gained from this discussion to infer the conditions under which SBRS might appear 
in solitonic models and in that context we study its appearance in baby Skyrme models.
Specifically, we shall show that SBRS emerges if the domain manifold of the model is
a two-sphere, while if the domain is $\mathbb{R}^2$,
SBRS does not occur. 

\subsection{SBRS from a dynamical point of view}
The onset of SBRS may be qualitatively understood as resulting from 
a competition between the static energy of a system and its moment of inertia.
To see this, 
let us consider a system described by a set of degrees of freedom $\phi$,
and assume that the dynamics of the system is governed by a Lagrangian which is invariant under spatial rotations.
When the system is static, its equilibrium configuration
is obtained by minimizing its static energy $E_{\textrm{static}}$ with respect to its degrees of freedom $\phi$
\begin{equation} \label{eq:Estat1}
\frac{\delta E}{\delta \phi}=0 \quad \textrm{where} \quad E=E_{\textrm{static}}(\phi) \,.
\end{equation}
Usually, if $E_{\textrm{static}}(\phi)$ does not include terms which manifestly break rotational
symmetry, the solution to (\ref{eq:Estat1}) is rotationally-symmetric
(with the exception of degenerate spontaneously-broken vacua, which are not of our concern 
here).
If the system rotates with a given angular momentum $\bJ=J \hat{z}$, its configuration is
naturally deformed. Assuming that
the Lagrangian of the system is quadratic in the time derivatives,
stable rotating configurations (if such exist) are obtained by minimizing its total
energy  $E_{J}$
\begin{equation} \label{eq:Erot1}
\frac{\delta E_{J}}{\delta \phi}=0 \quad \textrm{where} \quad E_{J}=
E_{\textrm{static}}(\phi)+\frac{J^2}{2 I(\phi)} \,, 
\end{equation}
where $I(\phi)$ is the ratio between the angular momentum of the system and its angular velocity  $\bomega=\omega \hat{z}$
(which for simplicity we assume is oriented in the direction of the angular momentum). 
$I(\phi)$ is the (scalar) moment of inertia of the system.
\par
The energy functional (\ref{eq:Erot1})  consists of two terms.
The first, $E_{\textrm{static}}$,
increases with the asymmetry. This is simply a manifestation of the 
minimal-energy configuration in the static case
being rotationally-symmetric. The second term
$J^2/2I$, having the moment of inertia in the denominator,
decreases with the asymmetry. 
At low values of angular momentum, the $E_{\textrm{static}}$ term 
dominates, and thus asymmetry is not energetically favorable, but as the value of angular momentum
increases, the second term becomes dominant, 
giving rise to a possible breaking of rotational symmetry.

\subsection{\label{sec:origBaby} SBRS in baby Skyrme models}
In what follows, we show
that the above mechanism of SBRS
is present in certain types of baby Skyrme models.
\par
As already discussed in previous sections, the static solutions of the baby Skyrme model (\ref{eq:BabyLag}) have  
rotationally-symmetric energy and charge distributions in the charge-one and charge-two sectors \cite{Old1}. 
The charge-one skyrmion has an energy peak at its center which drops down exponentially.
The energy distribution of the charge-two skyrmion has a ring-like peak around its center at some 
characteristic distance.
The rotating solutions of the model in $\mathbb{R}^2$ are also known \cite{Old2,rotBaby2}.
Rotation at low angular velocities slightly deforms
the skyrmion but it remains rotationally-symmetric.
For larger values of angular velocity, the rotationally-symmetric configuration becomes
unstable but in this case the skyrmion does not undergo symmetry breaking. 
Its stability is restored through a different mechanism, namely that of radiation. 
The skyrmion radiates out the excessive energy and angular momentum,
and as a result begins slowing down until it reaches equilibrium at some constant angular velocity,
its core remaining rotationally-symmetric. 
Moreover, if the Skyrme fields are 
restricted to a rotationally-symmetric (hedgehog) form,
the critical angular velocity above which the skyrmion radiates can be obtained analytically. 
It is simply the coefficient of the potential term $\omega_{\textrm{crit}}=\mu$
\cite{Old2}.
Numerical full-field simulations also show that the skyrmion
actually begins radiating well below $\omega_{\textrm{crit}}$,
as radiation itself may be non-rotationally-symmetric. 
The skyrmion's core, however, remains rotationally-symmetric for every angular velocity.
\par 
The stabilizing effect of the radiation on the solutions of the model
has lead us to believe that models in which radiation 
is somehow inhibited may turn out to be good candidates for the occurrence of SBRS.
In what follows, we study the baby Skyrme model on the two-sphere,
whose static solutions were presented in the previous section.
Within this model, energy and angular momentum are not allowed to escape 
to infinity through radiation,
and as a consequence, for high enough angular momentum the mechanism responsible for SBRS discussed in the previous section 
takes over, revealing solutions with spontaneously broken rotational symmetry.

\subsection{\label{sec:2sphere} The baby Skyrme model on the two-sphere}
In order to find the stable rotating solutions of the model,
we assume for simplicity that 
any stable solution would rotate
around the axis of angular momentum (which is taken to be the $z$ direction) with some angular velocity $\omega$.
The rotating solutions thus take the form \hbox{$\bphi(\theta,\varphi,t)=\bphi(\theta,\varphi- \omega t)$}. 
The energy functional to be minimized is
\bea \label{eq:Eomega}
E=  E_{\textrm{static}}+ \frac{J^2}{2 I} \,,
\eea
where $I$ is the ratio of the angular momentum of the skyrmion to its angular velocity, or its 
``moment of inertia'', given by
\bea
I = \frac1{4 \pi B}  \int \rmd \Omega \left((\partial_{\varphi} \bphi)^2 
+\kappa^2 (\partial_{\theta} \bphi \times \partial_{\varphi} \bphi)^2
\right) \,.
\eea
\subsection{Results}
In what follows we present the results obtained by the minimization scheme
applied to the rotating solutions of the model
in the charge-one and charge-two sectors, which as mentioned above are rotationally-symmetric.
For simplicity, we fix the parameter $\kappa$ at $\kappa^2= 0.01$ 
although other $\kappa$ values were tested as well,
yielding qualitatively similar solutions. 

\subsubsection{Rotating charge-one solutions} 
\begin{figure}[ht!] \begin{center}
\includegraphics[angle=0,scale=1,width=0.95\textwidth]{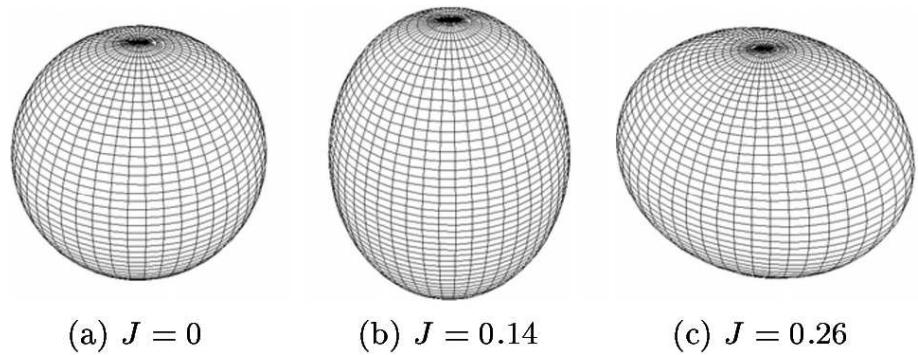}
\caption{\label{O3b1}Baby skyrmions on the two-sphere ($\kappa^2=0.01$): 
The charge distribution $\mathcal{B}(\theta, \varphi)$ of the charge-one skyrmion for different angular momenta.
In the figure, the vector 
$\mathcal{B}(\theta, \varphi)\vecr$
is plotted for the various $\theta$ and $\varphi$ values.$\hfill$}
\end{center}
\end{figure}

The rotating charge-one skyrmion has spherically-symmetric
energy and charge distributions
in the static limit (Fig. \ref{O3b1}a).
When rotated slowly, its symmetry is reduced to $O(2)$, 
with the axis of symmetry coinciding with the axis of rotation (Fig. \ref{O3b1}b).
At some critical value of angular momentum (which in the current settings is 
$J_{\textrm{crit}} \approx 0.2$), the axial symmetry 
is further broken, yielding an ellipsoidal
energy distribution with three unequal axes (Fig. \ref{O3b1}c). Any further increase in angular momentum
results in the elongation of the skyrmion in one horizontal direction and its shortening 
in the perpendicular one. The results are very similar to 
those of the rotating self-gravitating ellipsoid.

\newpage
\subsubsection{Rotating charge-two solutions}
\begin{figure}[htp!] \begin{center}
\includegraphics[angle=0,scale=1,width=0.9\textwidth]{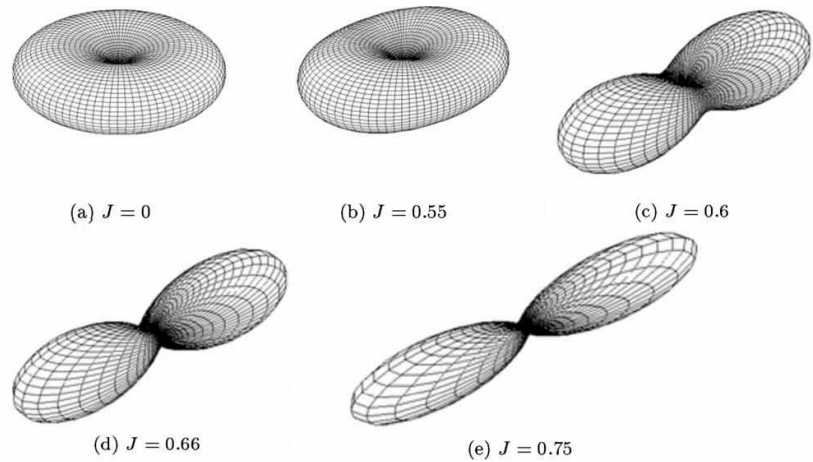}
\caption{\label{O3b2}Baby skyrmions on the two-sphere ($\kappa^2=0.01$): 
The charge distribution $\mathcal{B}(\theta, \varphi)$ of the charge-two skyrmion for different angular momenta.
In the figure, the vector 
$\mathcal{B}(\theta, \varphi) \vecr$
is plotted for the various $\theta$ and $\varphi$ values.$\hfill$}
\end{center} \end{figure}

SBRS is also observed in rotating charge-two skyrmions.
The static charge-two skyrmion has only axial symmetry (Fig. \ref{O3b2}a),
with its symmetry axis having no preferred direction. 
Nonzero angular momentum aligns the axis of symmetry with the axis of rotation.
For small values of angular momentum, the skyrmion is slightly deformed but remains axially-symmetric (Fig. \ref{O3b2}b).
Above $J_{\textrm{crit}} \approx 0.55$ however, its rotational 
symmetry is broken, and it starts splitting to its `constituent' 
charge-one skyrmions (Fig.  \ref{O3b2}c and \ref{O3b2}d). 
As the angular momentum is further increased, the splitting becomes more evident,
and the skyrmion assumes a string-like shape.  This is somewhat reminiscent of the
well-known elongation, familiar from high-spin hadrons which are also known to assume 
a string-like shape with the constituent quarks taking position at the ends of the string \cite{Nambu,Kang}.

\subsection{The rational map ansatz}
A somewhat more analytical analysis of this system may be achieved by the use 
of the rational maps approximation scheme \cite{Rmaps3DSk1}, which as
was shown earlier provides
quite accurate results for the static solutions of the model \cite{HK3}.

In its implementation here, we simplify matters even more and reduce the degrees of freedom
of the maps by a restriction only to 
those maps which exhibit the symmetries observed in the rotating full-field solutions.
This allows the isolation of those parameters which are the most critical for
the minimization of the energy functional.
\par
\begin{figure}[ht!] \begin{center}
\includegraphics[angle=0,scale=1,width=0.97\textwidth]{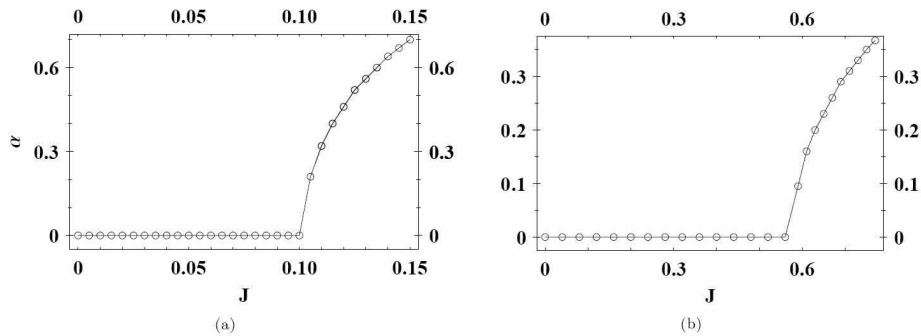}
\caption{\label{O3Rmap}Spontaneous breaking of rotational symmetry 
in the restricted rational maps approximation for the baby skyrmions on the two-sphere:
the parameter $\alpha$ as a function of the angular momentum $J$,
for the charge-one (top) and the charge-two (bottom) solutions.
The lines are to guide the eye.$\hfill$}
\end{center} \end{figure}

As  shown in Fig.~\ref{O3b1}, the charge and energy densities of the charge-one skyrmion
exhibit progressively lower symmetries as $J$ is increased. The static solution has an $O(3)$ symmetry,
while the slowly-rotating solution has an $O(2)$ symmetry. Above a certain critical $J$, the
$O(2)$ symmetry is further broken and only an ellipsoidal symmetry survives. 
Rational maps of degree one, however, cannot
produce charge densities which have all the discrete symmetries of an ellipsoid with three unequal axes. 
Nonetheless, approximate solutions with only a reflection symmetry through the $xy$ plane 
(the plane perpendicular to the axis of rotation) and a reflection through 
one horizontal axis may be generated by the following one-parametric family of rational maps
\bea \label{eq:RmapB1}
R(z)=\frac{\cos \alpha}{z+ \sin \alpha} \,, 
\eea
which has the charge density
\bea
\mathcal{B}(\theta, \varphi)=\left( \frac{\cos \alpha}{1 +\sin \alpha \sin \theta \cos \varphi} \right)^2 \,. 
\eea
Here, $\alpha \in [-\pi, \pi]$ is the parameter of the map, with $\alpha=0$ corresponding to a 
spherically-symmetric solution and a non-zero value of $\alpha$ corresponding to  a nonrotationally-symmetric solution.
Results of a numerical minimization of the energy functional (\ref{eq:Eomega}) for fields constructed 
from (\ref{eq:RmapB1})
for different values of angular momentum $J$ are shown in
Fig.~\ref{O3Rmap}a.
While for angular momentum less than $J_{\textrm{crit}} \approx 0.1$,
$\alpha=0$ minimizes the energy functional (a spherically-symmetric solution),
above this critical value bifurcation occurs and $\alpha=0$ is no longer a minimum;
the  rotational symmetry of the charge-one skyrmion is broken and it becomes nonrotationally-symmetric.
\par
A similar analysis of the charge-two rotating solution yields the one-parametric map
\bea \label{eq:RmapB2}
R(z)=\frac{\sin \alpha +z^2 \cos \alpha}{\cos \alpha +z^2 \sin \alpha} \,,
\eea
with corresponding charge density
\bea
\mathcal{B}(\theta, \varphi)= \left( \frac{2 \cos 2 \alpha \sin \theta}
{2 + \sin^2 \theta  (\sin 2 \alpha \cos 2 \varphi-1)} \right)^2 \,. 
\eea
In this case, $\alpha=0$ corresponds to a torodial configuration, and a 
non-zero value of $\alpha$
yields solutions very similar to those shown in Fig.~\ref{O3b2},
having the proper discrete symmetries.
The results in this case are summarized in Fig.~\ref{O3Rmap}b,
indicating that above $J_{\textrm{crit}} \approx 0.57$ the minimal energy configuration
is no longer axially-symmetric. 

\par
The discrepancies in the critical angular momenta $J_{\textrm{crit}}$
between the full-field method ($0.2$ for charge-one and $0.55$ for charge-two) 
and the rational maps scheme ($0.1$ for charge-one and $0.57$ for charge-two) 
are of course expected,
as in the latter method, the solutions have only one degree of freedom.
Nonetheless, the qualitative similarity in the behavior of the solutions 
in both cases is strong. 

\subsection{Further remarks}
We have seen that SBRS appears not only in rotating classical-mechanical systems
but also in the baby Skyrme model on the two-sphere.
We have argued that this is so because the phenomenon originates from general principles, 
and hence it is a universal one. 
\par
The results presented above may, at least
to some extent, also be linked to recent advances in the understanding
the non-sphericity of 
excited nucleons with of large orbital momentum.  
Non-spherical deformation of the
nucleon shape
is now a focus of considerable interest, both 
experimental \cite{exp1,exp2} and theoretical \cite{Miller1,Miller2,
Miller3}.
As skyrmions are
known to provide a good qualitative description of many nucleon properties, 
the results presented here may provide some corroboration to
recent results on this subject (e.g., ~\cite{Miller3}), although a more detailed
analysis of this analogy is in order.

\section*{Acknowledgments}
This work was supported in part by a grant from the Israel Science
Foundation administered by the Israel Academy of Sciences and Humanities.

\begin{appendix}
\section{Obtaining Baby Skyrmion Solutions --- The Relaxation Method}

As a large part of the studies presented above is based on 
numerically obtaining the baby skyrmion configurations, in the following
we describe the relaxation method that was used 
to obtain the solutions. 

The multi-solitons of the baby Skyrme model
are those field configurations which minimize the static energy
functional within each topological sector. 
The energy functional is given by
\bea \label{eq:O3energyApp}
 E=\int \rmd^2 x \, \left(
\frac1{2} (\partial_x \bphi \cdot \partial_x \bphi 
+ \partial_y \bphi \cdot \partial_y
+\frac{\kappa^2}{2} (\partial_x \bphi \times \partial_y \bphi)^2
+U(\phi_3)
\right) \;.
\eea

As already noted, the baby Skyrme model is a nonintegrable system, so
in general, explicit analytical solutions to
its Euler-Lagrange equations are nearly impossible to find.
Hence, one must resort to numerical techniques. 

Generally speaking, there are two main approaches to finding the baby skyrmion solutions numerically.
One approach is to employ standard techniques to numerically solve the Euler-Lagrange 
equations which follow from the energy functional (\ref{eq:O3energyApp}).
The other approach -- the one taken here -- 
is to utilize relaxation methods to minimize
the energy of the skyrmion within any desired 
topological sector. In what follows, we describe in some detail
the relaxation method we have used all throughout this research.
This method is based on the work of Hale, Schwindt and Weidig \cite{SkyrmeSA}.
We assume for simplicity that the base space is descretized to a rectangular grid.
The implementation of this method in the case of curved spaces or for a polar grid is
straightforward.
\par
The relaxation method begins by defining a grid with $N^2$ points,
where at each point a field triplet $\bphi(x_m,y_n)$ is defined. 
All measurable quantities such as energy density or charge density are calculated at the centers 
of the grid squares, using the following expressions for the 
numerical derivatives, also evaluated at these points:
\bea
 &&\frac{\partial\bphi}{\partial x}\Big|_{(x_{m+\frac{1}{2}},y_{n+\frac{1}{2}})} 
= \frac1{\Delta x} \times  \\\nonumber  \Bigg( 
  && {\Big \langle} \frac{\bphi(x_{m+1},y_{n}) + \bphi(x_{m+1},y_{n+1})}{2}  {\Big \rangle}_{\rm normed}
-{\Big \langle}\frac{\bphi(x_{m},y_{n}) + \bphi(x_{m},y_{n+1})}{2}  {\Big \rangle}_{\rm normed} 
 \Bigg)\,,
\eea
with the $y$-derivatives analogously defined, and the ``normed'' subscript indicates that the averaged fields
are normalized to one.  If the field itself has to calculated at that center of a grid square, we use
the prescription
\bea
&&\bphi(x_{m+\frac{1}{2}},y_{n+\frac{1}{2}})\\\nonumber
&=& {\Big \langle} \frac{1}{4} \left( 
\bphi(x_{m},y_{n}) + \bphi(x_{m},y_{n+1}) +\bphi(x_{m+1},y_{n}) + \bphi(x_{m+1},y_{n+1})
\right) {\Big \rangle}_{\rm normed} \;. 
\eea
The basic updating mechanism of the relaxation process consists of the following two steps:
A point $(x_m,y_n)$ on the grid is chosen at random, along 
with one of the three components of the field $\bphi(x_m,y_n)$.
The chosen component is then shifted by a value $\delta_{\phi}$ chosen uniformly from the segment $[-\Delta_{\phi},\Delta_{\phi}]$
where $\Delta_{\phi}=0.1$ initially. The field triplet is then scaled
and the change in energy is calculated.
If the energy decreases, the modification of the field is accepted
and otherwise it is discarded.

The relaxation process, through which
the energy of the baby skyrmion is minimized, is as follows:
\begin{enumerate}
\item
Initialize the field triplet $\bphi$
to a rotationally--symmetric configuration  
\bea \label{eq:initConfig}
\bphi_{\textrm{initial}}=(\sin f(r) \cos B \theta,\sin f(r) \sin B \theta,\cos f(r)) \quad.
\eea
In our setup, we have chosen the profile function $f(r)$ to be $f(r)=\pi \exp(-r)$,
$r$ and $\theta$ being the usual polar coordinates.
\item  \label{it:3}
Perform the basic updating mechanism for $M \times N^2$ times (we took $M =100$),
and then calculate the average rate of acceptance. If it is smaller than $5 \%$, 
decrease $\Delta_{\phi}$ by half. 
\item
Repeat step (\ref{it:3})  until $\Delta_{\phi}<10^{-9}$,
meaning no further decrease in energy is observed.
\end{enumerate}

This procedure was found to work very well in practice,
and its accuracy and validity were verified by comparison of our results to known ones. 
There is however one undesired feature to this minimization scheme, which we note here: it
can get stuck at a local minimum.
This problem can be resolved by using the ``simulated annealing'' algorithm \cite{SA1,SA2},
which in fact has been successfully implemented before,
in obtaining the minimal energy configurations of 
three dimensional skyrmions \cite{SkyrmeSA}. 
The algorithm is comprised of repeated applications of a Metropolis algorithm 
with a gradually decreasing temperature,
based on the fact that when a physical system is slowly cooled down,
reaching thermal equilibrium at each temperature, it will end up in its ground state. 
This algorithm, however, is much more expensive in terms of computer time. 
We therefore employed it only in part,
just as a check on our method, which corresponds to a Metropolis algorithm algorithm at zero temperature.
We found no apparent changes in the results.
\end{appendix}

\bibliographystyle{ws-rv-van}
\bibliography{ws-rv-sample}

\printindex                         
\end{document}